\documentstyle[prl,aps,preprint]{revtex}

\begin{document}
\renewcommand{\theequation}{\arabic{section}.\arabic{equation}}

\widetext
\draft
\title{
Topological Defects on Fluctuating Surfaces: \\
General Properties and the Kosterlitz-Thouless Transition
}
\author{Jeong-Man Park and T. C. Lubensky}
\address{
Department of Physics, University of Pennsylvania, \\
Philadelphia, PA 19104
}
\maketitle

\begin{abstract}
We investigate the Kosterlitz-Thouless transition for hexatic order on a
free fluctuating membrane and derive both a Coulomb gas and a sine-Gordon
Hamiltonian to describe it.  
The Coulomb-gas Hamiltonian includes charge densities arising from
disclinations and from Gaussian curvature.  There is an interaction
coupling the difference between these two densities, whose strength is
determined by the hexatic rigidity, and an interaction coupling Gaussian
curvature densities arising from the Liouville Hamiltonian resulting from
the imposition of a covariant cutoff.  
In the sine-Gordon Hamiltonian,
there is a linear coupling between a scalar field and the Gaussian
curvature. We discuss gauge-invariant correlation function for hexatic
order and the dielectric constant of the Coulomb gas.  
We also derive renormalization group recursion relations that
predict a transition with decreasing bending rigidity $\kappa$.
\end{abstract}

\section{Introduction}
\label{sec:1}
Bilayer fluid membranes\cite{SM-Jerus,reviews}
spontaneously self assemble when aliphatic
molecules are dissolved in water at a sufficiently high concentration.
At high temperature, these membranes have no internal order and can be
modeled as fluctuating structureless surfaces characterized by a bare
bending rigidity $\kappa$.  The bending rigidity is length-scale dependent
and becomes zero at the persistence length $\xi_p = a e^{4 \pi \kappa/3T}$
where $a$ is a molecular length and $T$ is the temperature.  At length
scales less than of order
$\xi_p$, the membrane is flat; at longer length scales, it
is crumpled.
\par
\begin{figure}
\centerline{\psfig{figure=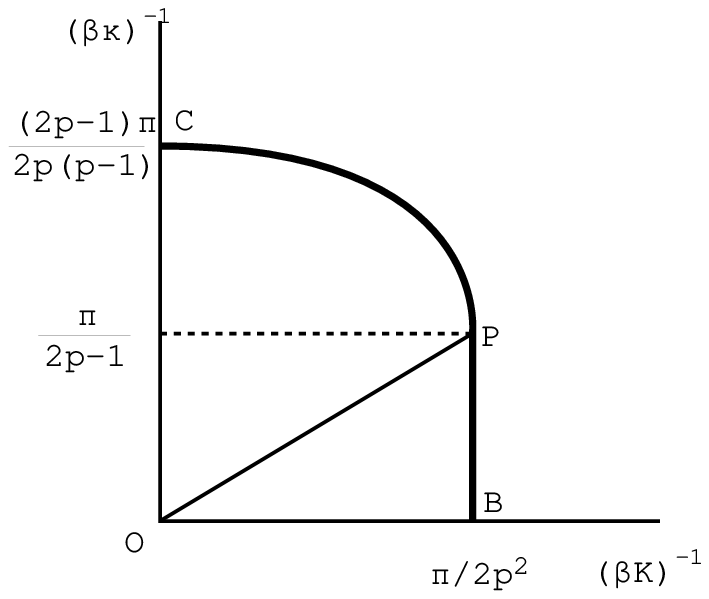}}
\caption{Schematic phase diagram in the $(\beta K)^{-1}$-$(\beta \kappa
)^{-1}$ plane adapted from \protect\onlinecite{GuitKardar90-1}. 
The vertical line $BP$ is
the Kosterlitz-Thouless disclination melting line of a flat membrane.  
The curved line $PC$ is an
estimate of the crumpling critical line based upon 
a comparison of the energy and entropy of
a single positive disclination in a buckled membrane. 
The line $OP$ is the mechanical 
buckling instability line of a membrane with a single 
positive disclination.  Above this line the zero-temperature 
membrane is buckled.  Decreasing $K$ or increasing $T$ 
at fixed $\kappa$ in 
the vicinity of $P$ leads to a disclination melting 
transition to the crumpled phase.  
On decreasing $\kappa$ at fixed $K$, 
the crumpled phase can only be reached by crossing the crumpling line. }
\label{GTfig}
\end{figure}

A flat rigid membrane can have quasi-long-range (QLR) hexatic order 
\cite{nel-halp} at low
temperature and undergo a Kosterlitz-Thouless (KT)  disclination
unbinding transition \cite{KT,nel-domb,minnhagen} 
to a disordered high-temperature phase.  A fluctuating 
membrane can also have QLR hexatic order \cite{NelPel}.  
Hexatic order stiffens
the bending rigidity so that, rather than scaling to zero at long length
scales, it approaches a constant times the hexatic rigidity 
$K$ \cite{David87-1,GuitKardar90-1}.  
The hexatic membrane is thus more rigid than a fluid
membrane, and it is said to be ``crinkled" rather than crumpled.  A
fluctuating hexatic membrane can undergo a KT transition
from the crinkled to the crumpled state.  
Reference \onlinecite{GuitKardar90-1} discussed two possible mechanisms
for the crinkled-to-crumpled transition: disclination melting and
crumpling.  The latter mechanism is analogous to that producing the
flat-to-crumpled transition in tethered membranes\cite{tethered} and is
argued to be associated with the buckling instability\cite{seung-nel}
of a membrane with a single disclination.  Figure \ref{GTfig} shows a
schematic phase-flow diagram in the $(\beta K)^{-1}$-$(\beta\kappa )^{-1}$ 
plane ($\beta = 1/T$) for a $2d$
membrane embedded in three dimensions adapted from Ref.
\onlinecite{GuitKardar90-1}.  The vertical line at $(\beta K)^{-1} = \pi
/72$ is the Kosterlitz-Thouless disclination unbinding line of a flat
membrane.  The curved line joining the vertical line at $(\beta
\kappa)^{-1} = \pi/11 $ is an estimate of the crumpling transition obtained
by equating the energy of a single positive disclination in a buckled
membrane to its entropy.  Thus, the crumpling transition in this estimate
is a Kosterlitz-Thouless transition in a buckled membrane.  This
schematic phase diagram describes qualitatively features that are in
agreement with simple physical reasoning: for large $\kappa$, there
should be a disclination melting to the crumpled phase as
temperature is increased, and at fixed $K$, there should be a transition
to the crumpled phase as $\kappa$ is decreased.  
It, however, has features that are either unexpected or unexplained.  The
discontinuous change in curvature where the melting and crumpling lines
join is surely an artifact.  It suggests that the physics of the
crinkled-to-crumpled transition produced by decreasing $K$ at fixed
$\kappa$ and by decreasing $\kappa$ at fixed $K$ are totally different,
the latter being associated with the buckling instability of a membrane
with a single disclination. It leaves unanswered whether the buckling
instability line for the zero temperature membrane has any significance
for a membrane in thermal equilibrium, which is allowed to choose
disclination configurations to minimize its free energy and thereby to
reject highly energetic configurations with an excess of positive or
negative disclinations.
\par
\begin{figure}
\centerline{\psfig{figure=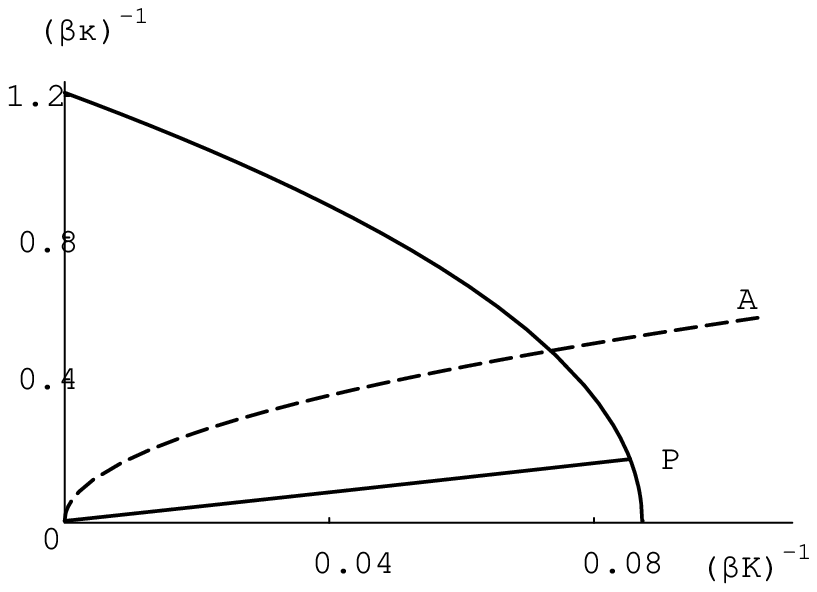}}
\caption{Phase diagram for hexatic membranes ($p=6$)
in the $(\beta K)^{-1}$-$(\beta \kappa )^{-1}$ plane
obtained from Eqs.\ (\protect\ref{RG1}), (\protect\ref{RG2}), and
(\protect\ref{kappa1}).  
The dark curved line  
[ $(\beta K)^{-1} + (3/32 \pi)(\beta \kappa )^{-2} = \pi/72 = 0.0436$] 
is the critical line separating the crinkled from 
the crumpled phase.  The line $OP$ is the
crinkled line with 
$4(\beta \kappa)^{-1} = (\beta K)^{-1} + (3/32 \pi) (\beta \kappa )^{-2}]$.
$P$ is the crinkled-to-crumpled fixed point.  
Unlike Fig. \protect\ref{GTfig}, the critical line has curvature at $P$.
The dashed curve $OA$ is the curve
$2 \pi (\beta K)^{-1} =  (\beta \kappa )^{-2}$.  The calculations 
in this paper are 
approximately valid in the region below this line where 
$(\beta \kappa)^{-1} <1$ and $\beta
K < 2 \pi ( \beta \kappa )^2$.  The crinkled-to-crumpled 
transition occurs via disclination
melting in the vicinity of $P$ when either $\kappa$ or $K$ 
is decreased or when temperature
is increased.  Though our calculations do not apply 
above the curve $OA$, it plausible that
the crumpled-to-crinkled transition occurs via disclination 
melting for all $\kappa$ and $K$ except at $K= \infty$, i.e., that
the only effect of higher order terms in $(\beta \kappa)^{-1}$ 
and $\beta K/(\beta \kappa
)^2$ is to change the shape of the critical line.  
Alternatively, there could be some other
phase boundary above the curve $OA$ separating melting 
from some kind of crumpling transition.}

\label{hexphased}
\end{figure}

In this paper, we present a detailed analysis of the low-temperature
crinkled-to-crumpled transition in hexatic membranes.  
Our approach treats both disclinations and Gaussian curvature in the same
real-space renormalization procedure and allows us to obtain recursion
relations for $\kappa$, $K$ and the disclination fugacity $y$.
Previous treatments used momentum space renormalization procedures
to calculate the recursion relation for $K$ and did not actually
provide a complete set of recursion relations for $\kappa$, $K$ and $y$
nor a prescription for doing so.
Our procedure shows that thermally induced shape fluctuations cause
a $\kappa$-dependent reduction in $K$, missed in previous calculation,
that leads to the phase-flow diagram shown in Fig.~\ref{hexphased}.
The vertical line in Fig.~\ref{GTfig} is now curved, and the mechanism
for the crinkled-to-crumpled phase transition in the vicinity of $P$,
the termination of the crinkled line, is the same for decreasing both 
$\kappa$ and $K$.
Our renormalization equations allow us to study the fluid phase in the
vicinity of the transition point $P$ and to show that near $P$ the
persistance length is $\xi_{KT}e^{4 \pi {\overline\kappa} /3 T}$ 
rather than $ae^{4\pi\kappa/3T}$, where 
$\xi_{KT} = a \exp (b / |T - T_{KT} |^{1/2} )$ is the KT correlation length
(with $b$ a constant), and ${\overline \kappa}$ is the bending rigidity at
length scale $\xi_{KT}$.
\par
Our calculations are strictly speaking restricted
to $(\beta \kappa)^{-1} < 1$, $(\beta K)^{-1} <1$ and $(\beta K)/(\beta
\kappa )^2 = T K/\kappa^2 < 1$, i.e., to the region below the curve $OA$
in Fig. \ref{hexphased}.  We cannot, therefore, make any definitive
statement about the interesting $K\rightarrow \infty$ limit, which should
be related to the crumpling of tethered membranes.  However, we believe
that the phase diagram shown in Fig. \ref{hexphased} makes physical sense
beyond the region of validity of our calculations. In this scenario, the
transition from the crinkled to the crumpled phase would be controlled by
the fixed point $P$ in Fig. \ref{hexphased} for all $K^{-1} >0$ and
$\kappa^{-1}>0$.  The transition at $K= \infty$ would, however, be
controlled by another fixed point.
\par
We begin in Sec.II with a review of how to describe
tangent plane order on fluctuating surfaces.  We pay particular attention
to correlation functions of orientational order.  In order to compare
tangent plane vectors at two different points on the surface, it is
necessary to parallel transport one of the vectors along some path $\Gamma$
to the position of the other vector.  When Gaussian curvature is nonzero,
the direction of a parallel transported vector depends on $\Gamma$ even
when there are no disclinations present.  Physical correlation functions
are invariant with respect to local coordinate transformations and for a
particular membrane shape and distribution of disclinations depend on
$\Gamma$.  When correlation functions are averaged over shape and position
of disclinations, the dependence on $\Gamma$ vanishes.
\par
In Sec.III, we discuss various models for hexatic membranes.  We begin
with the Hamiltonian for hexatic membranes expressed in terms of the
orientational angle $\theta$ and a height variable.  We then transform this
model into a Coulomb gas model in which ``charge" density arises both from
disclinations and from Gaussian curvature.  There are two Coulomb-like
terms in this Hamiltonian: one which is zero when the local disclinations
density equals the local Gaussian curvature and one, arising from the
imposition of a covariant cutoff\cite{polyakov}, 
that couples Gaussian curvature to
Gaussian curvature.  Finally, we transform the Coulomb-gas Hamiltonian 
to a sine-Gordon Hamiltonian with a term coupling the sine-Gordon field
$\phi$ to the Gaussian curvature with an imaginary coefficient.  The latter
term is analogous to the dilaton coupling of string theory \cite{dilaton}.
\par
In Sec.IV, we relate the dielectric constant and hexatic rigidity to
correlation functions of the disclination-Gaussian curvature density.  We
show in particular that the renormalized hexatic rigidity appearing in 
orientational correlation function is the same as that calculated from the
free energy.  We also calculate the charge-density correlation functions.
\par
Finally in Sec.V, we derive renormalization group recursion relations for
$K$, $\kappa$, and the disclination fugacity $y$.  These equations show
that height fluctuations renormalize the hexatic rigidity in the absence of
disclinations. This renormalized rigidity then flows
under renormalization in the presence of disclination
in exactly the same way as the rigidity of a flat membrane.
If the initial height renormalized rigidity is less than the 
critical rigidity for a flat membrane, there is no rigid phase.  Thus,
height fluctuations can destroy the crinkled phase.
\par
In Sec. VI, we review results and discuss some unanswered questions.
In particular, we address the effect of asymmetry in the energies of
positive and negative disclinations\cite{NEL} on our results. 
\par
The focus of this paper is on the nature of tangent-plane order and the
Kosterlitz-Thouless transition on fluctuating surfaces.  It will not
present a complete derivation of the RG equations used in Sec.V because
they involve some two loop graphs, which should be treated in a
sophisticated regularization procedure.  In a companion paper, we will
derive the complete recursion relations from the sine-Gordon model using a
real space regularization procedure generalized from that used by Amit et
al. \cite{amit} to treat the flat space problem. 
\setcounter{equation}{0}
\section{Differential Geometry and Tangent-Plane Order}
\label{sec:2}
We are concerned with the nature of tangent-plane order on fluctuating
membranes. In this section, we will review relevant concepts in differential
geometry, mostly to establish notation. 
We will also discuss how to describe long-range order (or lack thereof)
on a metric with fluctuating curvature.
\subsection{Differential Geometry of a Plane}
Points \cite{choquet,dubro2,David-Jerus} on a 
two-dimensional surface embedded in three-dimensional
Euclidean space are specified by a three-dimensional vector
${\bf R} ({\bf u})$ with components $R_{i}({\bf u}), \; i=1,2,3,$ 
as a function of
a two-dimensional parameter ${\bf u} = (u^{1},u^{2})$. 
Covariant tangent-plane vectors are then defined as
\begin{equation}
{\bf t}_{\alpha} = \partial_{\alpha}{\bf R}, \;\;\;\;\; \alpha=1,2,
\end{equation}
where $\partial_{\alpha}=\partial/\partial u^{\alpha}$.
We will use Greek letters $\alpha,\beta,\gamma,...$ to denote components
of covariant and contravariant tangent-plane tensors and Roman letters
$i,j,k,...$ to denote components of vectors and tensors in
Euclidean space.
The metric tensor is
\begin{equation}
g_{\alpha\beta} = {\bf t}_{\alpha}\cdot{\bf t}_{\beta}.
\end{equation}
Its inverse $g^{\alpha\beta}$ satisfying
\begin{equation}
g^{\alpha\beta}g_{\beta\gamma}=\delta^{\alpha}_{\gamma}   
\end{equation}
allows us to define contravariant tangent-plane vectors ${\bf t}^{\alpha} =
g^{\alpha\beta}{\bf t}_{\beta}$ satisfying 
${\bf t}^{\alpha}\cdot{\bf t}_{\beta} =
\delta^{\alpha}_{\beta}$.
Any vector ${\bf V}$ in the tangent plane can be expressed as
${\bf V} = V^{\alpha}{\bf t}_{\alpha} = V_{\alpha}{\bf t}^{\alpha}$ where
$V_{\alpha} = {\bf t}_{\alpha}\cdot{\bf V}$ and 
$V^{\alpha} = {\bf t}^{\alpha}\cdot{\bf V} =
g^{\alpha\beta}V_{\beta}$ are, respectively, the covariant and contravariant
components of ${\bf V}$.
A unit normal ${\bf N}$ to the surface can be constructed from 
${\bf t}_{1}$ and ${\bf t}_{2}$ :
\begin{equation}
{\bf N} = \frac{{\bf t}_{1}\times{\bf t}_{2}}{|{\bf t}_{1}\times{\bf t}_{2}|}.
\end{equation}
The curvature tensor is then
\begin{equation}
K_{\alpha\beta} = {\bf N}\cdot\partial_{\alpha}\partial_{\beta}{\bf R}.
\end{equation}
From the curvature tensor, one can construct
the mean curvature,
\begin{equation}
\frac{1}{2} H = \frac{1}{2}K^{\alpha}_{\alpha} = 
\frac{1}{2}\left(\frac{1}{R_{1}}+\frac{1}{R_{2}}\right),
\end{equation}
and the Gaussian curvature,
\begin{equation}
S = \det K^{\alpha}_{\beta} = \frac{1}{R_{1}}\cdot\frac{1}{R_{2}},
\end{equation}
where $R_{1}$ and $R_{2}$ are the principal radii of curvature at the point of
the surface in question.
The integral of the Gaussian curvature is a topological invariant,
\begin{equation}
\int d^{2}u\sqrt{g} S = 4\pi(1-\eta) = 2\pi\chi,
\label{eq:EulerC}
\end{equation}
where $\eta$ is the number of handles and $\chi = 2(1-\eta)$ is the Euler
characteristic.
In the Monge gauge, 
${\bf u} = (x,y)$ and ${\bf R}({\bf u}) = ({\bf u},h({\bf u}))$, and
the metric tensor $g_{\alpha\beta}$ is written as
\begin{equation}
g_{\alpha\beta} = \partial_{\alpha}{\bf R}\cdot\partial_{\beta}{\bf R}
= \left(  \begin{array}{cc}
          1+(\partial_{x}h)^{2} & \partial_{x}h\partial_{y}h  \\
          \partial_{x}h\partial_{y}h  & 1+(\partial_{x}h)^{2} 
v          \end{array}
  \right),
\end{equation}
and the curvature tensor $K_{\alpha\beta}$ is
\begin{equation}
K_{\alpha\beta} = {\bf N}\cdot D_{\alpha}D_{\beta}{\bf R}
= \frac{-1}{\sqrt{1+(\nabla h)^{2}}} \left( \begin{array}{cc}
             \partial_{x}\partial_{x}h & \partial_{x}\partial_{y}h   \\
             \partial_{y}\partial_{x}h & \partial_{y}\partial_{y}h 
             \end{array}  \right),
\end{equation}
where $(\nabla h)^{2} = (\partial_{x}h)^{2} + (\partial_{y}h)^{2}$.

The anti-symmetric tensor $\gamma_{\alpha\beta}$ will be particularly useful
in what follows. It is defined via
\begin{eqnarray}
\gamma_{\alpha\beta} & = & {\bf N}\cdot({\bf t}_{\alpha}\times{\bf t}_{\beta})
                           \nonumber  \\
                     & = & \frac{g_{\alpha 1}g_{\beta 2} - 
        g_{\alpha 2}g_{\beta 1}}{|{\bf t}_{1}\times{\bf t}_{2}|}  \nonumber  \\
                     & = & \sqrt{g}\epsilon_{\alpha\beta},
\end{eqnarray}
where $g = \det g_{\alpha\beta}$ and $\epsilon_{\alpha\beta}$ is the 
anti-symmetric tensor with $\epsilon_{12} = -\epsilon_{21} = 1$.
The contravariant tensor 
\begin{equation}
\gamma^{\alpha\beta} = {\bf N}\cdot({\bf t}^{\alpha}\times{\bf t}^{\beta})
\end{equation}
equals $\epsilon_{\alpha\beta}/\sqrt{g}$ and satisfies 
$\gamma^{\alpha\beta}\gamma_{\beta\alpha'} = -\delta^{\alpha}_{\alpha'}$.
Finally the mixed tensor
\begin{equation}
{\gamma^{\alpha}}_{\beta} = g^{\alpha\alpha'}\gamma_{\alpha'\beta}
\label{eq:gam-rot}
\end{equation}
rotates vectors by $\pi/2$ since $V_{\alpha}{\gamma^{\alpha}}_{\beta}V^{\beta}
= \gamma_{\alpha\beta}V^{\alpha}V^{\beta} = 0$ and ${\gamma^{\alpha}}_{\beta}
V^{\beta}{\gamma_{\alpha}}^{\beta'}V_{\beta'} = V^{\alpha}V_{\alpha}$.
\par
We are interested primarily in order in the tangent plane of unit
(or fixed) magnitude. For this, it is useful to introduce orthonormal
tangent-plane basis vectors ${\bf e}_{1}$ and ${\bf e}_{2}$ satisfying
\begin{equation}
{\bf e}_{a}\cdot{\bf e}_{b} = \delta_{ab},  \;\; {\bf N}\cdot{\bf e}_{a} = 0.
\end{equation}
A tangent vector ${\bf V}$ can be expressed in the basis 
$\{ {\bf e}_{1}, {\bf e}_{2} \}$
as well as that defined by the covariant or contravariant vectors :
${\bf V} = V_{a}{\bf e}_{a}$ where $V_{a} = 
{\bf e}_{a}\cdot{\bf V}$. We will use Roman subscripts $a,b,c,...$ to
denote vector and tensor components with respect to the local orthonormal
basis. Covariant derivatives are derivatives projected into the tangent plane.
Components of the covariant derivative of a vector ${\bf V}$ relative to the
orthonormal basis are
\begin{eqnarray}
D_{\alpha}V_{a} & \equiv & {\bf e}_{a}\cdot(\partial_{\alpha}{\bf V}) 
                     = \partial_{\alpha} V_{a} +
                     {\bf e}_{a}\cdot\partial_{\alpha}{\bf e}_{b} V_{b}  
                     \nonumber  \\
   & = & \partial_{\alpha} V_{a} + \epsilon_{ab}A_{\alpha}V_{b},
\end{eqnarray}
where
\begin{equation}
A_{\alpha} = {\bf e}_{1}\cdot\partial_{\alpha}{\bf e}_{2}
\end{equation}
is the spin-connection whose curl is the Gaussian curvature :
\begin{equation}
\gamma^{\alpha\beta}\partial_{\alpha} A_{\beta} = S.
\label{eq:gaussian}
\end{equation}
We will also find it useful to use a circular basis defined by the vectors
\begin{equation}
\mbox{\boldmath $\epsilon$}_{\pm} = 
\frac{1}{\sqrt{2}}({\bf e}_{1} \pm i{\bf e}_{2}) = 
\mbox{\boldmath $\epsilon$}^{*}_{\mp},
\end{equation}
satisfying $\mbox{\boldmath $\epsilon$}_{a}\cdot
\mbox{\boldmath $\epsilon$}^{*}_{b} = \delta_{ab}$ with $a,b=\pm$.
In this basis, ${\bf V} = \tilde{V}_{a}\mbox{\boldmath $\epsilon$}^{*}_{a}$, 
and 
the covariant derivative,
\begin{eqnarray}
D_{\alpha}\tilde{V}_{\pm} & = & \mbox{\boldmath $\epsilon$}_{\pm}\cdot
       \partial_{\alpha}{\bf V} =
       \partial_{\alpha}\tilde{V}_{\pm} + 
       \mbox{\boldmath $\epsilon$}_{\pm}\cdot\partial_{\alpha}
       \mbox{\boldmath $\epsilon$}^{*}_{a}V_{a}
       \nonumber  \\
       & = & \partial_{\alpha}\tilde{V}_{\pm} \mp iA_{\alpha}\tilde{V}_{\pm} 
       \nonumber  \\
       & = & (\partial_{\alpha} \mp iA_{\alpha})\tilde{V}_{\pm},
\label{eq:cov-circ}
\end{eqnarray}
has a particularly simple form.
\subsection{Vector and Tensor Order}
A vector order parameter ${\bf S}$ that is restricted to lie in the tangent
plane of a surface can be written as $S^{\alpha}{\bf t}_{\alpha}$ or
$S_{a}{\bf e}_{a}$. If ${\bf S}$ is a unit length vector, it is conveniently
expressed in terms of its angle in the local orthonormal basis as
\begin{equation}
{\bf S} = \cos\theta {\bf e}_{1} + \sin\theta {\bf e}_{2} = S_{a}{\bf e}_{a},
\end{equation}
where $S_{1}=\cos\theta$ and $S_{2}=\sin\theta$.
The unit vector ${\bf S}$ can thus be written in the circular basis as
\begin{equation}
{\bf S} = \sqrt{2} \; {\rm Re} \; \mbox{\boldmath $\Psi$}
\end{equation}
where
\begin{equation}
\mbox{\boldmath $\Psi$} = \psi\mbox{\boldmath $\epsilon$}_{-} 
\equiv e^{i\theta}\mbox{\boldmath $\epsilon$}_{-}
\end{equation}
is a complex order parameter with unit length 
$\mbox{\boldmath $\Psi$}\cdot\mbox{\boldmath $\Psi$}^{*} = 1$.
\par
We now turn to tensor tangent-plane order. The simplest nontrivial
tensor order parameter is a symmetric-traceless tensor ${\bf Q}$ 
with Cartesian components
\begin{equation}
Q_{ij} = Q_{ab}e_{ai}e_{bj}.
\end{equation}
The traceless constraint implies $Q_{ii}=Q_{aa}=0$. ${\bf Q}$ is a 
uniaxial tensor
with a principal axis lying along a unit vector ${\bf N} = n_{a}{\bf e}_{a}$ in
the tangent plane. If we require that ${\bf Q}$ have a fixed magnitude 
defined by
${\rm Tr} {\bf Q}^{2} = 1$, then we can write
\begin{equation}
Q_{ab} = \sqrt{2}(n_{a}n_{b} - \frac{1}{2}\delta_{ab})
\end{equation}
or
\begin{equation}
Q_{ij}=\frac{1}{\sqrt{2}}[\cos 2\theta (e_{1i}e_{1j}-e_{2i}e_{2j})
+ \sin 2\theta (e_{1i}e_{2j}+e_{2i}e_{1j})].
\end{equation}
This tensor, like the vector ${\bf S}$, can be expressed in terms of the real
part of a complex tensor. Introduce the direct product tensor
\begin{equation}
\mbox{\boldmath $\Psi$}_{2} = 
\mbox{\boldmath $\Psi$}\otimes\mbox{\boldmath $\Psi$} = 
\psi_{2}\mbox{\boldmath $\epsilon$}_{-}\otimes\mbox{\boldmath $\epsilon$}_{-} 
\equiv
e^{2i\theta}\mbox{\boldmath $\epsilon$}_{-}
\otimes\mbox{\boldmath $\epsilon$}_{-}
\end{equation}
with components $\Psi_{ij} = e^{2i\theta}\epsilon_{-i}\epsilon_{-j}$.
Then
\begin{equation}
{\bf Q} = \sqrt{2} \; {\rm Re} \; \mbox{\boldmath $\Psi$}_{2}.
\end{equation}
Generalization of this construction to higher order tensors is straightforward.
Let
\begin{equation}
\mbox{\boldmath $\Psi$}_{p} = 
\mbox{\boldmath $\Psi$}\otimes\cdots\otimes\mbox{\boldmath $\Psi$} = 
\psi_{p}\mbox{\boldmath $\epsilon$}_{-}\otimes\cdots
\otimes\mbox{\boldmath $\epsilon$}_{-}
\label{eq:Ppsi}
\end{equation}
with $\psi_{p} = e^{ip\theta}$ be a $p$th rank complex tangent-plane tensor.
If we define the inner product of two tensors by
\begin{equation}
\mbox{\boldmath $\Psi$}_{p}\cdot\mbox{\boldmath $\Phi$}^{*}_{p} = 
\mbox{\boldmath $\Psi$}_{i_{1}\cdots i_{p}}
\mbox{\boldmath $\Phi$}^{*}_{i_{1}\cdots i_{p}}, \;\; 
\mbox{\boldmath $\Psi$}_{p}\cdot\mbox{\boldmath $\Psi$}^{*}_{p} = 
|\mbox{\boldmath $\Psi$}_{p}|^{2} = 1,
\end{equation}
then a real tensor, symmetric under interchanges of all indices and traceless
with respect to all pairs of indices, is 
\begin{equation}
{\bf Q}_{p} = \sqrt{2} \; {\rm Re} \; \mbox{\boldmath $\Psi$}_{p}.
\label{eq:patic}
\end{equation}
The tensor $\mbox{\boldmath $\Psi$}_{1} = \mbox{\boldmath $\Psi$}$ describes 
vector order such as is present in
the smectic-$C$ phase in which the long axes of molecules comprising 
the membrane 
tilt relative to the normal ${\bf N}$. $\mbox{\boldmath $\Psi$}_{2}$ 
describes tangent-plane nematic 
order with inversion symmetry. $\mbox{\boldmath $\Psi$}_{6}$ describes 
hexatic order.
$\mbox{\boldmath $\Psi$}_{4}$ would describe ``4-atic'' order, etc. Thus 
$\mbox{\boldmath $\Psi$}_{p}$ for general $p$ describes what we call 
``$p$-atic'' order.
\subsection{Tangent-plane Correlations and Parallel Transport}
In flat space, the basis vectors ${\bf e}_{1,2}$ or 
$\mbox{\boldmath $\epsilon$}_{\pm}$ are 
independent of spatial coordinate. Information about the existence of
long-range order and about $p$-atic order parameter correlations is
contained in the correlation function 
$\langle \mbox{\boldmath $\Psi$}_{p}({\bf x})\cdot
\mbox{\boldmath $\Psi$}^{*}_{p}({\bf x}') \rangle = 
\langle e^{-ip(\theta({\bf x}')-\theta({\bf x}))}
\rangle$. 
Basis vectors at different points on curved surfaces are not identical
(or even in the same plane) and the simple dot product of vectors at 
different points does not carry information about tangent-plane order.
In order to compare order parameters 
$\mbox{\boldmath $\Psi$}_{p}({\bf u})$ and 
$\mbox{\boldmath $\Psi$}^{*}_{p}({\bf u}')$ at two different points ${\bf u}$ 
and ${\bf u}'$, we need
to parallel transport the order parameter at ${\bf u}$ along some path 
$\Gamma$ to ${\bf u}'$. The parallel transported tensor 
$\mbox{\boldmath $\Psi$}^{\Gamma}_{p}
({\bf u},{\bf u}')$ is now in the tangent plane at ${\bf u}'$ like 
$\mbox{\boldmath $\Psi$}^{*}_{p}
({\bf u}')$ and we can take its dot product with 
$\mbox{\boldmath $\Psi$}^{*}_{p}({\bf u}')$.
Thus the correlation function,
\begin{equation}
G_{p}({\bf u},{\bf u}') = \langle \mbox{\boldmath $\Psi$}_{p}({\bf u}) : 
\mbox{\boldmath $\Psi$}^{*}_{p}({\bf u}') \rangle
\equiv \langle \mbox{\boldmath $\Psi$}^{\Gamma}_{p}({\bf u},{\bf u}')\cdot
\mbox{\boldmath $\Psi$}^{*}_{p}({\bf u}') \rangle,
\label{eq:gp-corr1}
\end{equation}
where the ``$:$'' means parallel transported inner product, is what is 
appropriate
for describing correlations in tangent-plane order. Note that this correlation
function depends on the path $\Gamma$ when the membrane has nonzero
Gaussian curvature. We will find, however, that it becomes independent of
$\Gamma$ on nearly flat membranes after averaging over height fluctuations.
\par
The parallel transported order parameter can be expressed in terms of the
spin-connection. Let ${\bf V}({\bf u}) = V_{a}({\bf u}){\bf e}_{a}({\bf u})$ 
be a tangent 
vector at ${\bf u}$ and let ${\bf V}^{\parallel}_{a}({\bf u}+\delta{\bf u}) = 
V^{\parallel}_{a}({\bf u}+\delta{\bf u})
{\bf e}_{a}({\bf u}+\delta{\bf u})$ be the vector ${\bf V}({\bf u})$ parallel 
transported to
a nearby point ${\bf u}+\delta{\bf u}$. 
Then by definition 
$V^{\parallel}_{a}({\bf u}+\delta{\bf u}) = V_{a}({\bf u}) + 
\delta V_{a}({\bf u})$
with
\begin{equation}
\delta V_{a}({\bf u}) = -\mbox{\boldmath $\epsilon$}_{ab} 
A_{\alpha}V_{b}({\bf u})\delta u^{\alpha}.
\end{equation}
Alternatively in terms of the circular basis,
\begin{equation}
\delta V_{\pm}({\bf u}) = 
\pm iV_{\pm}({\bf u})A_{\alpha}({\bf u})\delta u^{\alpha},
\end{equation}
or
\begin{equation}
\tilde{V}^{\Gamma}_{\pm} ({\bf u},{\bf u}') = e^{\pm i\int^{{\bf u}'}_{{\bf u}}
A_{\alpha}({\bf u})du^{\alpha}} \tilde{V}_{\pm}({\bf u}).
\end{equation}
Applying this result to the vector 
$\mbox{\boldmath $\Psi$} = \psi\mbox{\boldmath $\epsilon$}_{-}$, we obtain
\begin{equation}
\mbox{\boldmath $\Psi$}^{\Gamma}({\bf u},{\bf u}') = 
e^{i\int^{{\bf u}'}_{{\bf u}} A_{\alpha}({\bf u})du^{\alpha}}
\mbox{\boldmath $\Psi$}({\bf u}),
\end{equation}
or
\begin{equation}
\mbox{\boldmath $\Psi$}^{\Gamma}({\bf u},{\bf u}') = 
e^{i\int^{{\bf u}'}_{{\bf u}} A_{\alpha}({\bf u})du^{\alpha}}
e^{i\theta({\bf u})}\mbox{\boldmath $\epsilon$}_{-}({\bf u}).
\end{equation}
Thus we have 
\begin{equation}
G_{1}({\bf u},{\bf u}') = \left\langle 
e^{-i(\theta({\bf u}')-\theta({\bf u})-\int^{{\bf u}'}_{{\bf u}}
A_{\alpha}({\bf u})du^{\alpha})} \right\rangle.
\end{equation}
This function is invariant under changes in the local coordinate system
({\it i.e.} under nonlocal rotations of the vectors 
${\bf e}_{1}$ and ${\bf e}_{2}$).
The generalization to $p$-atic order is straight forward since by construction
$\mbox{\boldmath $\Psi$}_{p}({\bf u}) : 
\mbox{\boldmath $\Psi$}^{*}_{p}({\bf u}') = 
(\mbox{\boldmath $\Psi$}({\bf u}):\mbox{\boldmath $\Psi$}^{*}({\bf u}'))^{p}$ 
so that
\begin{equation}
G_{p}({\bf u},{\bf u}') = \left\langle 
e^{-ip(\theta({\bf u}')-\theta({\bf u})-\int^{{\bf u}'}_{{\bf u}}
A_{\alpha}({\bf u})du^{\alpha})} \right\rangle.  
\label{eq:gp-corr2}
\end{equation}
Again, this is invariant under coordinate system changes.
\setcounter{equation}{0}
\section{Model Hamiltonians}
\label{sec:3}
In this section, we will derive various equivalent representations 
for the Hamiltonian and associated partition function describing $p$-atic
order on a fluctuating surface. 
\subsection{Fluid Membranes}
It is now well established that the long wavelength properties of a fluid 
membrane are well-described by the Helfrich-Canham Hamiltonian
${\cal H}_{\rm HC}$ \cite{Hel-Natur,Can-Bio,SM-Jerus}, which
can be expressed as a sum of three terms,
\begin{equation}
{\cal H}_{\rm HC} = {\cal H}_{\kappa} + {\cal H}_{\rm G} + {\cal H}_{\sigma}.
\end{equation}
The first term is the mean curvature energy,
\begin{eqnarray}
{\cal H}_{\kappa} & = & \frac{1}{2}\kappa\int d^{2}u\sqrt{g} H^{2}
                        \nonumber  \\
                  & = & \frac{1}{2}\kappa\int d^2 u\sqrt{g}
\left( \nabla\cdot\frac{\nabla h}{1+(\nabla h)^{2}} \right)^{2}
\label{non-lin-cur}
\end{eqnarray}
where $H=K^{\alpha}_{\alpha}$, and the second form is valid for
the Monge gauge.
The second term is the Gaussian
curvature energy,
\begin{equation}
{\cal H}_{\rm G} = \frac{1}{2}\kappa_{G}\int d^{2}u\sqrt{g} S,
\end{equation}
where $S = \det K^{\alpha}_{\beta}$. This term is a topological 
invariant depending only on the genus of the surfaces.
Since we will consider surfaces of fixed genus, we will drop this term.
Finally
\begin{equation}
{\cal H}_{\sigma} = \sigma\int d^{2}u\sqrt{g}
\end{equation}
is the surface tension energy. We are mostly interested in free membranes
for which the renormalized surface tension obtained by differentiating
the total free energy ${\cal F}$ with respect to the total surface area
${\cal A}$ $(\sigma_{R} = \partial {\cal F}/\partial {\cal A})$, is zero.
Since there are entropic contributions to $\sigma_{R}$ as well as contributions
vfrom internal order, the value of the bare surface tension $\sigma$ will
have to be adjusted to keep $\sigma_{R}$ zero.
In what follows, we will ignore ${\cal H}_{\sigma}$ with the understanding
that it is really present if we want to keep track of how $\sigma_{R}$
actually becomes zero.
\par
The partition function for a fluid membrane,
\begin{equation}
{\cal Z}_{\rm Fl} = \int {\cal D}{\bf R}({\bf u}) e^{-\beta{\cal H}_{\rm HC}}
\end{equation}
is obtained by integrating over all physically 
distinct realizations of the surface, which is specified by the vector
${\bf R}({\bf u})$. 
This means we have to specify a gauge or parametrization for
${\bf R}({\bf u})$.
Thus the measure ${\cal D}{\bf R}({\bf u})$ for distinct surface configurations
contains a Fadeev-Popov determinant \cite{popov}. 
In addition, it in general contains
a factor to correct for the fact that different surface configurations
arising from a fixed parametrization surface can have different areas
\cite{Cai-Lub}.
\subsection{$p$-atic Order}
As discussed in the preceding section, the complex tangent-plane tensor
$\mbox{\boldmath $\Psi$}_{p}$ distinguishes a fluid membrane from one with 
$p$-atic
tangent-plane order. Since $\mbox{\boldmath $\Psi$}_{p}$ has a fixed 
magnitude and there are
no external fields aligning $\mbox{\boldmath $\Psi$}_{p}$ along 
a particular direction, the
lowest nontrivial contribution to the energy associated with 
$\mbox{\boldmath $\Psi$}_{p}$
arises from its gradients,
\begin{equation}
{\cal H}_{\theta} = \frac{1}{2}K_{p}\int d^{2}u\sqrt{g} g^{\alpha\beta}
D_{\alpha}\mbox{\boldmath $\Psi$}^{*}_{p}\cdot 
D_{\beta}\mbox{\boldmath $\Psi$}_{p},
\end{equation}
where $D_{\alpha}$ is a covariant derivative.
Using Eq.~(\ref{eq:cov-circ}) and Eq.~(\ref{eq:Ppsi}), it is straight forward
to show
\begin{equation}
D_{\alpha}\mbox{\boldmath $\Psi$}_{p} = 
ip(\partial_{\alpha}\theta - A_{\alpha}) \mbox{\boldmath $\Psi$}_{p}
\end{equation}
and
\begin{equation}
{\cal H}_{\theta} = \frac{1}{2}K\int d^{2}u\sqrt{g} g^{\alpha\beta}
(\partial_{\alpha}\theta - A_{\alpha})(\partial_{\beta}\theta - A_{\beta}),
\label{eq:patic-ener}
\end{equation}
where we set $K = K_{p}p^{2}$. 
Eq.~(\ref{eq:patic-ener}) is the simplest energy arising from $p$-atic order.
It is rotationally isotropic, and it describes correctly the lowest order
gradient contribution to the energy for all $p \geq 3$.
For $p=1$ or $p=2$, however, this energy should have anisotropic contributions
\cite{pel-prost,nel-powers}.
In this paper, we will ignore these anisotropies.
\par
The $p$-atic order parameter $\mbox{\boldmath $\Psi$}_{p}$ can have 
disclinations of strength
$ q= 2\pi(k/p)$ where $k$ is an integer \cite{chai-lub}. 
A disclination at ${\bf u} = {\bf u}_{\nu}$ 
with 
strength $q_{\nu}$ gives rise to a singular contribution $\theta^{\rm sing}$
to $\theta$ satisfying 
\begin{equation}
\oint_{\Gamma} du^{\alpha}\partial_{\alpha}\theta^{\rm sing} = q_{\nu},
\end{equation}
where $\Gamma$ is a contour enclosing ${\bf u}_{\nu}$.
Thus, in general 
$\partial_{\alpha}\theta = \partial_{\alpha}\theta' + 
v_{\alpha}$ where $\theta'$ is
nonsingular, $v_{\alpha} = \partial_{\alpha}\theta^{\rm sing}$ and
\begin{equation}
\gamma^{\alpha\beta}\partial_{\alpha} v_{\beta} = n({\bf u}),
\label{eq:vortex}
\end{equation}
where
\begin{equation}
n({\bf u}) = \frac{1}{\sqrt{g}} \sum_{\nu} q_{\nu}
\delta({\bf u} - {\bf u}_{\nu})
\end{equation}
is the disclination density.
The vector $v_{\alpha}$ can always be chosen so that it is purely transverse,
so that $D_{\alpha}v^{\alpha} = 0$. 
In the $p$-atic Hamiltonian, $\partial_{\alpha}\theta$ always occurs 
in the combination
$\partial_{\alpha}\theta - A_{\alpha}$. 
The spin-connection $A_{\alpha}$ can and will
in general have both a longitudinal and a transverse component.
However, one can always redefine $\theta'$ to include the longitudinal part
of $A_{\alpha}$. This amounts to choosing locally rotated orthonormal vectors
${\bf e}_{1}({\bf u})$ and ${\bf e}_{2}({\bf u})$ so that 
$D_{\alpha}A^{\alpha} = 0$.
Thus we may take both $v_{\alpha}$ and $A_{\alpha}$ to be transverse and
the $p$-atic Hamiltonian,
\begin{eqnarray}
{\cal H}_{\theta} & = & \frac{1}{2}K\int d^{2}u\sqrt{g}g^{\alpha\beta}
(\partial_{\alpha}\theta' + v_{\alpha} - A_{\alpha})
(\partial_{\beta}\theta' + v_{\beta} - A_{\beta})
          \\  \nonumber
   &  = & {\cal H}_{\parallel} + {\cal H}_{\perp},
\label{eq:h-theta}
\end{eqnarray}
can be decomposed into a regular longitudinal part,
\begin{equation}
{\cal H}_{\parallel} =  \frac{1}{2}K\int d^{2}u\sqrt{g}g^{\alpha\beta}
\partial_{\alpha}\theta'\partial_{\beta}\theta',
\end{equation}
and a transverse part,
\begin{equation}
{\cal H}_{\perp} =  \frac{1}{2}K\int d^{2}u\sqrt{g}g^{\alpha\beta}
(v_{\alpha} - A_{\alpha})(v_{\beta} - A_{\beta}),
\label{eq:h-perp}
\end{equation}
where it is understood that $D_{\alpha}(v^{\alpha}-A^{\alpha}) = 0$.
\par
It costs an energy $\epsilon_{c}(k)$ to create the core of a disclination
of strength $k$. (We assume for the moment that the core energies of the
positive and negative disclinations are the same.  See, however, Ref.\
\onlinecite{NEL} and the summary section.) 
Thus, partition sums should be weighted by a factor
$y_{k} = e^{-\beta\epsilon_{c}(k)}$ for each disclination of strength
$k$. Since $\epsilon_{c}(k) \sim k^{2}$, we may at low temperature
restrict our attention to configurations in which only configurations
of strength $\pm 1$ appear. Let $N_{\pm}$ be the number of disclinations
of strength $\pm 1$ and let ${\bf u}_{\nu^{\pm}}$ be the coordinate of the 
core of
the disclination with strength $\pm 1$ labeled by $\nu$. The $p$-atic membrane
partition function can then be written as
\begin{equation}
{\cal Z} (\kappa, K, y) = {\rm Tr}_{\rm v} y^{N}
\int{\cal D}{\bf R}\int{\cal D}\theta
e^{-\beta{\cal H}_{\kappa}}e^{-\beta{\cal H}_{\theta}},
\label{eq:partition}
\end{equation}
where $y=y_{1}$, and $N = N_{+} + N_{-}$.
${\cal H}_{\theta}$ depends on all of the disclination coordinates
${\bf u}_{\nu^{\pm}}$ where $\nu^{\pm}=1,2,\cdots,N_{\pm},$ and
\begin{equation}
{\rm Tr}_{\rm v} = \sum_{N_{+},N_{-}}\delta_{N_{+}+N_{-},p\chi}
\frac{1}{N_{+}!N_{-}!}
\prod_{\nu^{+}}\int\frac{d^{2}u_{\nu^{+}}}{a^{2}}
\prod_{\nu^{-}}\int\frac{d^{2}u_{\nu^{-}}}{a^{2}},
\label{eq:trace}
\end{equation}
where $a$ is a molecular length. The Kronecker factor 
$\delta_{N_{+}+N_{-},p\chi}$ in ${\rm Tr}_{\rm v}$ imposes the topological 
constraint \cite{topo-constraint} that the total disclination strength 
on a surface with Euler
characteristic $\chi$ be equal to $\chi$. Thus, on a sphere with $\chi = 2$,
$N_{+}+N_{-} = 2p$. If $p=6$ for example, there must be 12 more positive
than negative disclinations. On a nearly flat surface, the Coulomb interaction
effectively restricts $N_{+}$ to equal $N_{-}$, and we use Eq.~(\ref{eq:trace})
with $\chi = 0$ even though $\chi$ for an open surface is strictly
speaking equal to one \cite{nakahara}.
\subsection{The Coulomb Gas Model}
The $p$-atic model of Eq.~(\ref{eq:partition}) can easily be converted to a 
Coulomb gas model using
\begin{equation}
\gamma^{\alpha\beta}\partial_{\alpha}(v_{\beta}-A_{\beta}) = n - S \equiv \rho,
\label{eq:curl-vel}
\end{equation}
which follows from Eq.~(\ref{eq:gaussian}) and Eq.~(\ref{eq:vortex}).
The quantity $\rho = n-S$ is a ``charge'' density with contributions arising 
both from disclinations and Gaussian curvature.
Equation (\ref{eq:curl-vel}) implies
\begin{equation}
v_{\alpha}-A_{\alpha} = -{\gamma_{\alpha}}^{\beta}D_{\beta} 
\frac{1}{\Delta_{g}}\rho,
\label{eq:va-Aa}
\end{equation}
where $\Delta_{g}=D^{\alpha}D_{\alpha} = (1/\sqrt{g})\partial_{\alpha}\sqrt{g}
g^{\alpha\beta}\partial_{\beta}$ 
is the Laplacian on a surface with metric tensor 
$g_{\alpha\beta}$ acting on a scalar. 
Recall [Eq.~(\ref{eq:gam-rot})] that ${\gamma_{\alpha}}^{\beta}$ rotates
a vector by $\pi/2$ so that $v_{\alpha}-A_{\alpha}$ is perpendicular to
$D_{\beta}(-\Delta_{g})^{-1}\rho$ and is thus manifestly transverse.
Using Eq.~(\ref{eq:va-Aa}) in Eq.~(\ref{eq:h-perp}), we obtain
\begin{equation}
{\cal Z} = {\rm Tr}_{\rm v} y^{N}\int{\cal D}{\bf R}\int{\cal D}\theta' 
e^{-\beta
{\cal H}_{\kappa} -\beta{\cal H}_{\parallel} -\beta{\cal H}_{\rm c}},
\end{equation}
where
\begin{equation}
{\cal H}_{\rm c} = \frac{1}{2} K \int d^{2}u d^{2}u'
\sqrt{g}\rho({\bf u})\left( -\frac{1}{\Delta_{g}}\right)_{{\bf u}{\bf u}'}
\sqrt{g'}\rho({\bf u}')
\label{coulomb-rr}
\end{equation}
is the Coulomb Hamiltonian associated with the charge $\rho$.
The longitudinal variable $\theta'$ appears only quadratically in
${\cal H}_{\parallel}$ and the trace over $\theta'$ can be done directly:
\begin{equation}
\int {\cal D}\theta' e^{-\beta{\cal H}_{\parallel}} = 
e^{-\beta{\cal H}_{\rm L}},
\end{equation}
where
\begin{equation}
\beta{\cal H}_{\rm L} = \frac{1}{8\pi a^{2}}\int d^{2}u\sqrt{g} -
\frac{1}{24\pi}\int d^{2}u d^{2}u' \sqrt{g}S({\bf u}) 
\left(-\frac{1}{\Delta_{g}}\right)_{{\bf u}{\bf u}'} \sqrt{g'}S({\bf u}')
\end{equation}
is the Liouville action \cite{polyakov,David-Jerus} 
arising from the conformal anomaly.
The first term in this expression is proportional to the surface area and
can be incorporated into the surface tension energy ${\cal H}_{\sigma}$
by shifting the surface tension $\sigma$.
We assume, as discussed earlier, that the total surface tension is zero.
We will, therefore, ignore this term in what follows.
The second term can be viewed as a Coulomb interaction for Gaussian
curvature but with a negative coupling constant, {\it i.e.}, an imaginary
charge.
\par
The Coulomb gas partition function can thus be written
\begin{eqnarray}
{\cal Z} & = & {\rm Tr}_{\rm v} y^{N}\int{\cal D}{\bf R} 
e^{-\beta{\cal H}_{\kappa}
-\beta{\cal H}_{\rm L} -\beta{\cal H}_{\rm C}}  \nonumber  \\
   & = & {\rm Tr}_{\rm v} y^{N}\int{\cal D}{\bf R} e^{-\beta{\cal H}_{\kappa}
-\beta{\cal H}_{\rm CT}},
\label{eq:tr-hs}
\end{eqnarray}
where
\begin{eqnarray}
{\cal H}_{\rm CT} & = & \frac{1}{2} K \int d^{2}u d^{2}u'
\sqrt{g}\rho({\bf u})\left( -\frac{1}{\Delta_{g}}\right)_{{\bf u}{\bf u}'}
\sqrt{g'}\rho({\bf u}')  \nonumber  \\
   &   & -\frac{T}{24\pi}\int d^{2}u d^{2}u' \sqrt{g}S({\bf u}) 
\left(-\frac{1}{\Delta_{g}}\right)_{{\bf u}{\bf u}'} \sqrt{g'}S({\bf u}')  
\nonumber  \\
    & = & \frac{1}{2} K' \int d^{2}u d^{2}u' \sqrt{g}S({\bf u}) 
\left(-\frac{1}{\Delta_{g}}\right)_{{\bf u}{\bf u}'} \sqrt{g'}S({\bf u}')  
\nonumber   \\
   &   & + \frac{1}{2} K \int d^{2}u d^{2}u' \sqrt{g}n({\bf u}) 
\left(-\frac{1}{\Delta_{g}}\right)_{{\bf u}{\bf u}'} \sqrt{g'}n({\bf u}')  
\nonumber  \\
   &   & - K \int d^{2}u d^{2}u' \sqrt{g}n({\bf u}) 
\left(-\frac{1}{\Delta_{g}}\right)_{{\bf u}{\bf u}'} \sqrt{g'}S({\bf u}')  
\nonumber \\
   & & \equiv {\cal H}_{SS} + {\cal H}_{nn} + {\cal H}_{nS} 
\label{eq:tot-c-h}
\end{eqnarray}
is the ``total'' Coulomb Hamiltonian
in which there are two distinct charge densities, $\rho$ and $S$, or,
equivalently, $n$ and $S$. 
The coupling constant for Gaussian curvature interactions in the absence
of disclinations is
\begin{equation}
K' = K - \frac{T}{12\pi}.
\end{equation}
This is the shifted coupling arising from the Liouville term discussed in 
Ref. \cite{David87-1} and \cite{GuitKardar90-1}.
The coupling constant for disclinations and 
between disclinations and Gaussian curvature are not shifted by
the Liouville term and remain equal to $K$.
\subsection{The sine-Gordon Model}
The Coulomb gas model can be converted following standard procedures
into a sine-Gordon model. The first step is to carry out a 
Hubbard-Stratonovich transformation on $\beta{\cal H}_{\rm C}$:
\begin{equation}
e^{-\beta{\cal H}_{\rm C}} = e^{\beta{\cal H}_{\rm L}}
\int{\cal D}\phi e^{-1/(2\beta K)\int d^{2}u\sqrt{g}\partial^{\alpha}\phi
\partial_{\alpha}\phi} e^{i\int d^{2}u\sqrt{g}\rho\phi},
\end{equation}
where the Liouville factor $e^{\beta{\cal H}_{\rm L}}$ is needed to ensure
that $e^{-\beta{\cal H}_{\rm C}}$ be one when $\rho=0$. 
Inserting this in Eq.~(\ref{eq:tr-hs}), we obtain
\begin{equation}
{\cal Z} = {\rm Tr}_{\rm v} y^{N}\int{\cal D}{\bf R}{\cal D}\phi 
e^{-\beta{\cal H}_{\kappa}}e^{-\beta{\cal H}_{\phi}}
e^{i\int d^{2}u\sqrt{g}(n-S)\phi},
\end{equation}
where 
\begin{equation}
\beta{\cal H}_{\phi} = \frac{1}{2\beta K}\int d^{2}u\sqrt{g}
\partial^{\alpha}\phi\partial_{\alpha}\phi.
\end{equation}
The only dependence on disclinations is now in the term linear in $n$. Thus to carry
out ${\rm Tr}_{\rm v}$, we need only to evaluate
\begin{eqnarray}
   &   & {\rm Tr}_{\rm v} y^{N}e^{i\int d^{2}u\sqrt{g} n}   \nonumber   \\
   & = & \sum_{N_{+},N_{-}}\frac{1}{N_{+}!N_{-}!}\delta_{N_{+}-N_{-},p\chi}
y^{N_{+}+N_{-}} \left( \int\frac{d^{2}u\sqrt{g}}{a^{2}} 
e^{2\pi i\phi({\bf u})/p}\right)^{N_{+}}  
\left( \int\frac{d^{2}u\sqrt{g}}{a^{2}} 
e^{-2\pi i\phi({\bf u})/p}\right)^{N_{-}}   \nonumber  \\
   & = & \sum_{N_{+},N_{-}}\frac{1}{N_{+}!N_{-}!}
\int \frac{d\omega}{2\pi} e^{i\omega p\chi}
\left( y \int\frac{d^{2}u\sqrt{g}}{a^{2}} 
e^{i \{ 2\pi [\phi({\bf u})/p] -\omega \} }\right)^{N_{+}}    
\left( y \int\frac{d^{2}u\sqrt{g}}{a^{2}} 
e^{-i \{ 2\pi [\phi({\bf u})/p] -\omega \} }\right)^{N_{-}}     \nonumber  \\
   & = & \int \frac{d\omega}{2\pi} e^{i\omega p\chi}
e^{(2y/a^{2})\int d^{2}u\sqrt{g} \cos[2\pi(\phi/p)-\omega]}.
\end{eqnarray}
Thus
\begin{equation}
{\cal Z} = \int \frac{d\omega}{2\pi}\int{\cal D}\phi\int{\cal D}{\bf R}
               e^{-\beta{\cal H}_{\kappa}}e^{-\beta{\cal H}_{\phi}}
               e^{i\omega p\chi}  e^{(2y/a^{2})\int d^{2}u\sqrt{g}
               \cos[2\pi(\phi/p)-\omega]} e^{-i\int d^{2}u\sqrt{g}
               S\phi}.
\label{eq:tr-sg}
\end{equation}
We can now change variables, letting $\phi = (p/2\pi)(\phi'+\omega)$.
The term linear in the Gaussian curvature then becomes
\begin{equation}
-i\int d^{2}u\sqrt{g} S \frac{p}{2\pi}(\omega + \phi')
= -i\omega p\chi -i\frac{p}{2\pi}\int d^{2}u\sqrt{g} S\phi',
\end{equation}
where we used Eq.~(\ref{eq:EulerC}) relating $\chi$ to the integral of
the Gaussian curvature. Thus the $\omega$-dependent part of this term
exactly cancels the $i\omega p\chi$ term arising from the integral 
representation of the Kronecker delta. The integral over $\omega$
in Eq.~(\ref{eq:tr-sg}) is now trivial, and we obtain
\begin{equation}
{\cal Z} = \int{\cal D}\phi\int{\cal D}{\bf R} e^{-\beta{\cal H}_{\kappa}}
e^{-{\cal L}},
\label{eq:par-sg}
\end{equation}
where 
\begin{equation}
{\cal L} =  \frac{1}{2\beta K}\left(\frac{p}{2\pi}\right)^{2}
\int d^{2}u\sqrt{g} g^{\alpha\beta}\partial_{\alpha}\phi\partial_{\beta}\phi   
        -\frac{2y}{a^{2}}\int d^{2}u\sqrt{g} \cos\phi 
        -i\frac{p}{2\pi}\int d^{2}u\sqrt{g} S\phi
\end{equation}
is the sine-Gordon action on a fluctuating surface of arbitrary genus.
The first two terms of this action are the gradient and cosine energies
present in a flat space. The final term provides the principal coupling
between $\phi$ and fluctuations in the metric.
It is analogous to the dilaton coupling \cite{dilaton} of string theory 
though here
the coupling constant is imaginary rather than real.
Note that the Liouville action is not explicitly present in 
Eq.~(\ref{eq:par-sg}). The requirement that the cutoff $\phi$ be applied
in a covariant fashion is, however, still present.
Thus, if $g=0$, the integral over $\phi$ will yield the Liouville factor.
When $y$ is not zero, care must be taken to implement any cutoffs in
integrals in a covariant fashion.  
\setcounter{equation}{0}
\section{The Dielectric Constant and Hexatic Rigidity}
\label{sec:4}
In the preceding section, we derived several equivalent expressions
for the partition function of a fluctuating membrane with internal $p$-atic
order. In this section, we will derive an expression for the inverse
dielectric constant associated with the charge density $\rho$ and
show that it is equivalent to the renormalized $p$-atic rigidity controlling
the gauge-invariant $p$-atic correlation function of 
Eq.~(\ref{eq:gp-corr1}).
We will begin by reviewing the derivation of the longitudinal dielectric
constant in flat space. We will then derive the dielectric constant and
renormalized rigidities on fluctuating surfaces.
\subsection{Flat Space Dielectric Constant}
Consider a system with an internal charge density $\rho$ and fixed or
controllable external charge density $\rho_{\rm e}$.
To be concrete, $\rho_{\rm e}$ could be the free charge at the metal
electrodes of a capacitor.
The total charge density is $\rho_{\rm T} = \rho+\rho_{\rm e}$.
The electric field ${\bf {\cal E}}$ is determined by the total charge:
\begin{equation}
\nabla\cdot{\bf {\cal E}} = \alpha\rho_{\rm T}.
\end{equation}
Here $\alpha$ sets the units of charge (for example 
$\alpha = \epsilon_{0}^{-1}$ in {\it mks} units).
The displacement vector ${\bf {\cal D}} = \epsilon\alpha^{-1}{\bf {\cal E}}$, 
where $\epsilon$ is the dielectric constant, is controlled by the
external charge:
\begin{equation}
\nabla\cdot {\bf {\cal D}} = \nabla\cdot {\bf {\cal D}}_{\parallel} = 
\rho_{\rm e},
\end{equation}
where ${\bf {\cal D}}_{\parallel}$ is the longitudinal part of 
${\bf {\cal D}}$.
The electric potential can be introduced in the usual way: ${\bf {\cal E}} =
-\nabla\phi_{\rm T}$, 
${\bf {\cal D}}_{\parallel} = -\alpha^{-1}\nabla\phi_{\rm e}$
and $\phi_{\rm T} = \phi+\phi_{\rm e}$ where $-\nabla^{2}\phi = \alpha\rho$
and $-\nabla^{2}\phi_{\rm e} = \alpha\rho_{\rm e}$.
\par
The inverse longitudinal dielectric constant is obtained by expanding
the free energy to second order in 
${\bf {\cal D}}_{\parallel}=-\nabla\phi_{\rm e}$:
\begin{equation}
\Delta{\cal F} = \frac{1}{2}\alpha^{-1}\int d^{d}xd^{d}x'
\epsilon_{\parallel}^{-1}({\bf x},{\bf x}')
\nabla\phi_{\rm e}({\bf x})\cdot\nabla\phi_{\rm e}({\bf x}').
\end{equation}
A wave-number dependent dielectric constant can be obtained if
$\epsilon_{\parallel}^{-1}({\bf x},{\bf x}')$ is translationally invariant or 
by carrying the usual Maxwell averaging procedure.
The result is
\begin{equation}
\epsilon_{\parallel}^{-1}({\bf q}) = 
\frac{1}{q^{2}} \left(\frac{\alpha}{V}\right)
\int d^{d}xd^{d}x' e^{-i{\bf q}\cdot({\bf x}-{\bf x}')}
\frac{\delta^{2}{\cal F}}
{\delta\phi_{\rm e}({\bf x})\delta\phi_{\rm e}({\bf x}')} ,
\label{eq:diele-inv}
\end{equation}
where $V = \int d^dx$ is the volume of the system.
We can use this formula to relate $\epsilon_{\parallel}^{-1}({\bf q})$ to
correlations of the charge density.
The Coulomb Hamiltonian can be written as
\begin{eqnarray}
{\cal H}_{\rm C} & = & \frac{1}{2}\alpha\int d^{d}xd^{d}x' 
\rho_{\rm T}({\bf x})
\left(-\frac{1}{\nabla^{2}}\right)_{{\bf x}{\bf x}'} 
\rho_{\rm T}({\bf x}') \nonumber  \\
   & = & \frac{1}{2}\alpha^{-1}\int d^{d}x (\nabla\phi_{\rm e})^{2} +
\frac{1}{2}\int d^{d}x \rho\phi + \int d^{d}x \rho\phi_{\rm e}.
\end{eqnarray}
Expanding ${\cal F} = -T\ln e^{-\beta({\cal H}_{0}+{\cal H}_{\rm C})}$, where
${\cal H}_{0}$ is the non-Coulombic contribution to the Hamiltonian
to second order in $\nabla\phi_{\rm e}$, we obtain
\begin{equation}
\Delta{\cal F} = \frac{1}{2}\alpha^{-1}\int d^{d}x (\nabla\phi_{\rm e})^{2}
-\frac{1}{2}\beta\int d^{d}x \phi_{\rm e}({\bf x})\phi_{\rm e}({\bf x}')
\langle\rho({\bf x})\rho({\bf x}')\rangle
\end{equation}
and
\begin{equation}
\epsilon_{\parallel}^{-1}({\bf q}) = 1-\frac{\beta\alpha}{q^{2}} 
C_{\rho\rho}({\bf q}),
\label{eq:del-f}
\end{equation}
where
\begin{equation}
C_{\rho\rho}({\bf q}) = \frac{1}{V}\int d^{d}x d^{d}x' 
e^{-i{\bf q}\cdot({\bf x}-{\bf x}')}
\langle \rho({\bf x})\rho({\bf x}') \rangle
\label{eq:crr}
\end{equation}
is the density correlation function.
Eq.~(\ref{eq:del-f}) is the well-known expression \cite{p-martin}
relating $\epsilon^{-1}$ to
the charge susceptibility $\chi_{\rho\rho}=\beta C_{\rho\rho}$.
In what follows, we will use the notation introduced in Eq.~(\ref{eq:crr}) for
arbitrary variables, 
\begin{equation}
C_{AB}({\bf q})=\frac{1}{A_{B}}\int d^{2}x d^{2}x' 
e^{-i{\bf q}\cdot({\bf x}-{\bf x}')} \langle
A({\bf x})B({\bf x}') \rangle,
\label{eq:cab}
\end{equation}
where we have set $d=2$ and $V=A_{B}$.
\subsection{Dielectric Constant on a Fluctuating Membrane}
We can now determine the dielectric constant associated with the charge 
$\rho=n-S$ on a fluctuating membrane following exactly the procedures
outlined in the preceding subsection.
We impose external charges $\rho_{\rm e}$ to create a constant slowly
varying external potential $\phi_{\rm e}$.
These charges lead to a total charge density 
$\rho_{\rm T} = \rho+\rho_{\rm e}$.
They should not, however, change the Gaussian curvature because
otherwise they would change the Liouville energy. Thus $\rho_{\rm e}$ must
arise from disclinations which we can, for concreteness, place at the boundary
of the membrane like the charge on capacitor electrodes.
In analogy with flat space, we can introduce external and induced electric 
potentials $\phi_{\rm e}$ and $\phi$ satisfy $-\Delta_{g}\phi = \rho$ and
$-\Delta_{g}\phi_{\rm e} = \rho_{\rm e}$ with $\phi_{\rm T}=\phi+\phi_{\rm e}$
the total potential.
The Coulomb Hamiltonian can then be written as
\begin{eqnarray}
{\cal H}_{\rm C} & = & \frac{1}{2} K^{-1}\int d^{2}u\sqrt{g}
g^{\alpha\beta}\partial_{\alpha}\phi_{\rm e}\partial_{\beta}\phi_{\rm e}   
\nonumber  \\
   &  & + \frac{1}{2}\int d^{2}u\sqrt{g} \rho\phi + \int d^{2}u
\sqrt{g}\rho\phi_{\rm e}.
\end{eqnarray}
\par
We are interested in the long wavelength dielectric constant for our
presumed macroscopically flat (or nearly flat) membrane.
For this purpose it is most convenient to use the Monge gauge in which
${\bf u} = {\bf x} \equiv (x,y)$ measures position in the average plane
of the membrane. In this case $\phi_{\rm e}({\bf u}) \rightarrow 
\phi_{\rm e}({\bf x})$
is a slowly varying function of ${\bf x}$, which, in the limit of fixed charge
density at membrane boundaries, grows linearly with ${\bf x}$.
The long wavelength dielectric constant for our nearly flat membrane can
be obtained using Eq.~(\ref{eq:diele-inv}), where 
${\bf q}$ is the Fourier variable
dual to position ${\bf x}$, and Eq.~(\ref{eq:cab}) for ${\cal H}_{\rm C}$.
Using Eq.~(\ref{eq:cab}) and expanding ${\cal F}$ to second order in 
$\phi_{\rm e}$, we obtain
\begin{eqnarray}
\Delta{\cal F} & = & \frac{1}{2} K^{-1}\int d^{2}u \langle
\sqrt{g}g^{\alpha\beta} \partial_{\alpha}\phi_{\rm e}
\partial_{\beta}\phi_{\rm e} \rangle  \nonumber  \\
   &   & -\frac{1}{2}\beta\int d^{2}u \int d^{2}u' \langle
\sqrt{g}\rho({\bf x})\sqrt{g'}\rho({\bf x}') \rangle 
\phi_{\rm e}({\bf x})\phi_{\rm e}({\bf x}')  
\label{eq:del-free}
\end{eqnarray}
and
\begin{equation}
\epsilon_{\parallel}^{-1}({\bf q}) = \frac{q_{\alpha}q_{\beta}}{q^{2}}\langle
\sqrt{g}g^{\alpha\beta} \rangle - \frac{\beta K}{q^{2}}C_{\tilde{\rho}
\tilde{\rho}}({\bf q})
\end{equation}
is the charge density correlation function for
\begin{equation}
\tilde{\rho} = \sqrt{g}\rho
\end{equation}
rather than $\rho$.
$\tilde{\rho}$ is in fact the effective charge density for the coarse grained
flat surface (with metric tensor equal to unity) because the total charge 
in an area patch $d^{2}x$ should not change under coarse graining.
The quantity $\langle \sqrt{g}g^{\alpha\beta} \rangle$ is equal to 
$\delta^{\alpha\beta}$ as can be verified easily to lowest order in $T$ in
the Monge gauge or more generally using the conformal gauge 
\cite{conformalgauge,conformal}.
Thus we have
\begin{equation}
\epsilon_{\parallel}^{-1}({\bf q}) = 1-\frac{\beta K}{q^{2}} C_{\tilde{\rho}
\tilde{\rho}}({\bf q}).
\end{equation}
$C_{\tilde{\rho}\tilde{\rho}}({\bf q})$ is the charge density correlation
function evaluated in the presence of all interactions, including the
Coulomb like ${\cal H}_{\rm L}$.
We will return at the end of this section to its evaluation.
\subsection{Hexatic Rigidity}
In this section we will show that the renormalized hexatic rigidity $K_{R}$ is
equal to $\alpha\epsilon_{\parallel}^{-1}$.
To calculate $K_{R}$, we introduce a shift $\omega_{\alpha}$ in
$\partial_{\alpha}\theta'$ via $\partial_{\alpha}\theta' \rightarrow 
\partial_{\alpha}\theta'+\omega_{\alpha}$
in Eq.~(\ref{eq:h-theta}) and evaluate the free energy to second order in
$\omega_{\alpha}$.
Using Eq.~(\ref{eq:del-free}), we obtain
\begin{eqnarray}
\Delta{\cal F} & = & \frac{1}{2} K \int d^{2}u \langle
\sqrt{g}g^{\alpha\beta} \omega_{\alpha}\omega_{\beta} \rangle  \nonumber  \\
   &   & -\frac{1}{2}\beta K^{2}\int d^{2}u \int d^{2}u' \langle
\sqrt{g}D^{\alpha}\theta'\sqrt{g'}D^{\beta}\theta' \omega_{\alpha}
\omega_{\beta} \rangle    \nonumber  \\
   &   & -\frac{1}{2}\beta K^{2}\int d^{2}u \int d^{2}u' \langle
\sqrt{g}(v^{\alpha}-A^{\alpha})\sqrt{g'}(v^{\beta}-A^{\beta}) \omega_{\alpha}
\omega_{\beta} \rangle .
\label{eq:shift-free}
\end{eqnarray}
\par
As discussed previously, $\theta'$ decouples from other variables. 
For any given realization of the surface $\langle \theta\theta' \rangle =
(T/K)(-\Delta_{g}^{-1})$.
We can therefore combine the first two terms in Eq.~(\ref{eq:shift-free})
to obtain
\begin{eqnarray}
\Delta{\cal F} & = & \frac{1}{2} K \int d^{2}u d^{2}u' \langle
\sqrt{g} \omega_{\alpha} {\cal P}_{\perp}^{\alpha\beta}
\sqrt{g'} \omega_{\beta} \rangle  \nonumber  \\
   &   & -\frac{1}{2}\beta K^{2}\int d^{2}u \int d^{2}u' \langle
\sqrt{g}v^{\alpha}_{\rm T}\omega_{\alpha}\sqrt{g'}v^{\beta}_{\rm T}
\omega_{\beta} \rangle ,
\end{eqnarray}
where
\begin{equation}
{\cal P}_{\perp}^{\alpha\beta} = {\gamma^{\alpha}}_{\alpha'}
{\gamma^{\beta}}_{\beta'} \frac{D^{\alpha'}D^{\beta'}}{\Delta_{g}}
\frac{\delta(\tilde{u}-\tilde{u}')}{\sqrt{g}}
= \left( \delta^{\alpha\beta} - \frac{D^{\alpha}D^{\beta}}{\Delta_{g}} \right)
\frac{\delta(\tilde{u}-\tilde{u}')}{\sqrt{g}}
\end{equation}
is the transverse projection operator and
\begin{equation}
v^{\alpha}_{\rm T} = v^{\alpha} - A^{\alpha}
\end{equation}
is manifestly transverse from Eq.~(\ref{eq:va-Aa}).
Thus, as expected, $\Delta{\cal F}$ depends only on the transverse
part of $\omega_{\alpha}$.
\par
A longitudinal function such as $\partial_{\alpha}\phi_{\rm e}$ maintains 
its form under
coarse graining. A transverse function does not because the direction of a 
transverse function depends on the local value of the rotation matrix
${\gamma^{\alpha}}_{\beta}$.
Thus to coarse-grain our transverse function $\omega_{\alpha}({\bf x})$,
we set it equal to ${\gamma_{\alpha}}^{\beta}\partial_{\beta}\phi_{\rm e}$.
In the Monge gauge $\partial_{\beta}\phi_{\rm e}$ has the same value 
regardless of the
particular configuration of the membrane.
Then using ${\gamma^{\alpha}}_{\alpha'}{\gamma^{\alpha'}}_{\beta} =
-\delta^{\alpha}_{\beta}$, we obtain
\begin{eqnarray}
\Delta{\cal F} & = & \frac{1}{2} K \int d^{2}u \langle
\sqrt{g}g^{\alpha\beta} \partial_{\alpha}\phi_{\rm e}
\partial_{\beta}\phi_{\rm e} \rangle  \nonumber  \\
   &   & -\frac{1}{2}\beta K^{2}\int d^{2}u \int d^{2}u' \langle
\sqrt{g}\rho({\bf x})\sqrt{g'}\rho({\bf x}') \rangle 
\phi_{\rm e}({\bf x})\phi_{\rm e}({\bf x}')  .
\end{eqnarray}
Setting $\epsilon_{\alpha\beta}\partial_{\beta}\phi_{\rm e} = 
\bar{\omega}^{T}_{\beta}
(\delta_{\alpha\beta}-\partial_{\alpha}\partial_{\beta}/\partial^{2})$ 
where $\bar{\omega}_{\alpha}$
is the coarse grained $\omega_{\alpha}$, we find
\begin{eqnarray}
K_{R} & = & \frac{1}{A_{B}}\frac{\partial^{2}{\cal F}}
{\partial\bar{\omega}^{T}_{\alpha}\partial\bar{\omega}^{T}_{\alpha}}  
\nonumber  \\
   & = & \lim_{q\rightarrow 0}\left( K\frac{q_{\alpha}q_{\beta}}{q^{2}}
\langle \sqrt{g}g^{\alpha\beta} \rangle - \frac{\beta K^{2}}{q^{2}}
C_{\tilde{\rho}\tilde{\rho}}({\bf q}) \right) \nonumber  \\
   & = & K\epsilon_{\parallel}^{-1},
\end{eqnarray}
where $A_{B}$ is the base area and $\epsilon_{\parallel} = \epsilon_{\parallel}
(q=0)$.
Because $\int d^{2}x \tilde{\rho} = 0$, $C_{\tilde{\rho}\tilde{\rho}}({\bf q})$
tends to zero as $q \rightarrow 0$ in the ordered phase.
We can, therefore, define
\begin{eqnarray}
\lim_{q\rightarrow 0} \frac{1}{q^{2}} C_{\tilde{\rho}\tilde{\rho}}({\bf q})
  &  = & B   \nonumber   \\
  &  = & \lim_{q\rightarrow 0} \frac{1}{q^{2}} 
         (C_{\tilde{\rho}\tilde{\rho}}({\bf q})-
         C_{\tilde{\rho}\tilde{\rho}}(0))
         \nonumber  \\
  &  = & \lim_{q\rightarrow 0} \frac{1}{q^{2}} \frac{1}{A_{B}}
         \int d^{2}x d^{2}x' (e^{i{\bf q}\cdot({\bf x}-{\bf x}')}-1)
         \langle \tilde{\rho}({\bf x})\tilde{\rho}({\bf x}') \rangle  
         \nonumber  \\
  &  = & -\frac{1}{4} \int d^{2}x |{\bf x}|^{2} 
         \langle \tilde{\rho}({\bf x})\tilde{\rho}(0) \rangle  ,
\label{x2rr}
\end{eqnarray}
where in the final step we used the fact that 
$\langle \tilde{\rho}({\bf x})\tilde{\rho}({\bf x}') \rangle$ depends only on
${\bf x}-{\bf x}'$ in translationally invariant systems.
Thus
\begin{equation}
K_{R} = K - \beta K^{2} B.
\label{eq:k-ren}
\end{equation}
$C_{\tilde{\rho}\tilde{\rho}}({\bf q})$ is derived in Eq.~(\ref{eq:crr}).
\subsection{The $p$-atic correlation function}
We have argued that the renormalized rigidity is determined by the dielectric
constant associated with $\rho$ rather than that associated with some
other combination of the independent charge densities $n$ and $S$.
We will now show that it is indeed this rigidity that controls the long
distance properties of the $p$-atic correlation function $
G_{p}({\bf u},{\bf u}')$,
[Eqs.~(\ref{eq:gp-corr1}) and~(\ref{eq:gp-corr2})].
Our calculations follow closely those in Ref. \cite{Jose} for correlation
functions in flat space.
To keep things simple, we calculate only to lowest order in height
fluctuations in the Monge gauge.
\par
As discussed in Sec.II, the angle $\theta$ in Eq.~(\ref{eq:gp-corr2}) can be
broken up into a regular part and a singular part:
$\theta({\bf x})-\theta({\bf x}') = \theta'({\bf x})-\theta'({\bf x}')
+\int_{{\bf x}}^{{\bf x}'}ds^{\alpha}v_{\alpha}$.
In addition, we can always choose a gauge so that $A_{\alpha}$ is purely 
transverse.
In this case $\theta'$ decouples from other variables, and we have
\begin{equation}
G_{p}({\bf x}-{\bf x}') = \langle e^{-\frac{1}{2}p^{2}\langle (\theta'({\bf x})
-\theta'({\bf x}'))^{2}\rangle_{g}}e^{-ip\int_{{\bf x}}^{{\bf x}'}ds^{\alpha}
(v_{\alpha}-A_{\alpha})} \rangle_{{\bf R},{\rm v}},
\end{equation}
where $\langle (\theta'({\bf x})-\theta'({\bf x}'))^{2}\rangle_{g}$ is 
evaluated for a fixed metric tensor $g_{\alpha\beta}$ determined by 
${\bf R}$, and
\begin{equation}
\langle A\rangle_{{\bf R},{\rm v}} = {\cal Z}^{-1}{\rm Tr}_{\rm v}y^{N}
\int{\cal D}{\bf R} 
e^{-\beta{\cal H}_{\kappa}-\beta{\cal H}_{\rm CT}} A.
\end{equation}
To the order we are working, we can factorize $G_{p}({\bf x})$ into a 
``spin-wave'' part arising from $\theta'$ and a transverse part:
\begin{equation}
G_{p}({\bf x}) = G_{\rm sw}({\bf x}) G_{\rm v}({\bf x}),
\label{eq:gp-gen}
\end{equation}
where
\begin{equation} 
G_{\rm sw}({\bf x}) = e^{-\frac{1}{2}p^{2}
\langle \langle (\theta'({\bf x})-\theta'({\bf x}'))^{2}\rangle_{g}
\rangle}
\label{eq:doub-bra}
\end{equation}
and
\begin{equation}
G_{\rm v}({\bf x}) = e^{-\frac{1}{2}p^{2}\langle (\Delta\theta_{\rm v})^{2}
\rangle } = e^{-\frac{1}{2}p^{2} g_{\rm v}({\bf x})},
\label{eq:gv-gen}
\end{equation}
where $\Delta\theta_{\rm v} = \int_{{\bf x}}^{{\bf x}'} 
ds^{\alpha}(v_{\alpha}-A_{\alpha})$ and the double brackets in 
Eq.~(\ref{eq:doub-bra}) refers to an average over $\theta$ followed by an
average over ${\bf R}$.
In the Monge gauge, to the order we are working,
\begin{equation}
\langle (\theta'({\bf x})-\theta'({\bf x}'))^{2}\rangle_{g}
= \frac{2}{\beta K} {\cal G}({\bf x}-{\bf x}'),
\end{equation}
where
\begin{equation}
{\cal G}({\bf x}-{\bf x}') = \left( \frac{1}{\nabla^{2}} 
\right)_{{\bf x}{\bf x}'} =
\frac{1}{2\pi}\ln\frac{|{\bf x}-{\bf x}'|}{a}.
\end{equation}
Thus
\begin{equation}
G_{\rm sw}({\bf x}) = |{\bf x}|^{-p^{2}/2\pi\beta K}.
\label{eq:gsw-res}
\end{equation}
To evaluate $G_{\rm v}({\bf x})$, we use Eq.~(\ref{eq:va-Aa}) to lowest order 
in the Monge gauge to obtain
\begin{equation}
\Delta\theta_{\rm v} = \epsilon_{\alpha\beta}\int_{{\bf x}}^{{\bf x}'} 
d^{2}y\rho(y) f(y),
\end{equation}
where
\begin{equation}
f(y) = \int_{{\bf x}}^{{\bf x}'} ds^{\alpha}\epsilon_{\alpha\beta}
\partial^{s}_{\beta}{\cal G}({\bf s}-{\bf y}).
\end{equation}
Then using the Cauchy-Riemman relation,
\begin{eqnarray}
\epsilon_{\alpha\beta}\partial_{\beta}{\cal G}({\bf x}) & = & 
-\partial_{\alpha}{\cal G}_{\perp}({\bf x})
    \nonumber  \\
\epsilon_{\alpha\beta}\partial_{\beta}{\cal G}_{\perp}({\bf x}) & = & 
\partial_{\alpha}{\cal G}({\bf x}),
\end{eqnarray}
where ${\cal G}_{\perp}({\bf x}) = (1/2\pi)\tan (y/x)$, we find
\begin{equation}
f(y) = 
-({\cal G}_{\perp}({\bf x}'-{\bf y}) - {\cal G}_{\perp}({\bf x}-{\bf y}))
\label{eq:uy}
\end{equation}
and
\begin{eqnarray}
G_{\rm v}({\bf x}-{\bf x}') & = & \int d^{2}y d^{2}y' \langle 
\rho({\bf y})\rho({\bf y}') 
\rangle f({\bf y})f({\bf y}')  \nonumber  \\
   & = & -\frac{1}{2}\int d^{2}y d^{2}y' 
\langle \rho({\bf y})\rho({\bf y}')\rangle
(f({\bf y})-f({\bf y}'))^{2}.
\end{eqnarray}
Setting ${\bf y}={\bf R}+{\bf r}/2$ and ${\bf y}'={\bf R}-{\bf r}/2$, 
and expanding in ${\bf r}$, 
we obtain
\begin{equation}
G_{\rm v}({\bf x}-{\bf x}') = -\frac{1}{2}\int d^{2}r \langle 
\rho({\bf r})\rho(0) \rangle
r_{\alpha}r_{\beta} \int d^{2}R \partial_{\alpha} 
f({\bf R})\partial_{\beta} f({\bf R}).
\label{eq:gv-xx}
\end{equation} 
Then using Eqs.~(\ref{x2rr}),~(\ref{eq:uy}),~(\ref{eq:gv-xx}) 
and $\nabla_{R}^{2}{\cal G}
({\bf s}-{\bf R}) = \delta({\bf s}-{\bf R})$, we find
\begin{eqnarray}
G_{\rm v}({\bf x}-{\bf x}') & = & 2B [{\cal G}({\bf x}-{\bf x}')-{\cal G}(0)] 
\nonumber  \\
   & \sim & \frac{2B}{2\pi}\ln\frac{|{\bf x}-{\bf x}'|}{a},
\label{eq:gv-sim}
\end{eqnarray}
where we have treated ${\cal G}(0)$ as a non-divergent constant because
$({\bf x}-{\bf x}')$ is restricted to be greater than $a$.
\par
Combining Eqs.~(\ref{eq:gp-gen}),~(\ref{eq:gv-gen}),~(\ref{eq:gsw-res})
and~(\ref{eq:gv-sim}), we obtain
\begin{eqnarray}
G_{p} & = & |{\bf x}|^{-p^{2}/2\pi\beta K}|{\bf x}|^{-p^{2}B/2\pi}
    \nonumber  \\
   & = & |{\bf x}|^{-p^{2}/2\pi\beta K_{R}}
\end{eqnarray}
with
\begin{equation}
(\beta K_{R})^{-1} = (\beta K)^{-1} + B.
\end{equation}
This equation agrees with Eq.~(\ref{eq:k-ren}) to lowest order in $B$.
\subsection{Evaluation of the Dielectric Constant}
As we discussed at the beginning of this section, the correlation function
$C_{\tilde{\rho}\tilde{\rho}}({\bf q})$ must
be evaluated in the presence of
all interactions including that arising from the Liouville term.
In this subsection, we will derive a general expression for 
$C_{\tilde{\rho}\tilde{\rho}}({\bf q})$, which we will evaluate in the 
low-temperature, low-fugacity limit.
We will also discuss effective Coulomb interactions.
Both $\kappa$ and $K$ have units of energy.  Thus, both $(\beta
\kappa)^{-1}$ and $(\beta K)^{-1}$ are unitless measures of temperature.
In addition either of these quantities multiplied by any function of the
ratio $K/\kappa$ is also a measure of temperature.  In our
low-temperature expansion, we will require $A \beta K/(\beta \kappa )^2 =
A(T/\kappa)(K/\kappa)$, where $A$ smaller than $1/(2 \pi)$, as well as
$(\beta \kappa )^{-1}$ and $(\beta K)^{-1}$ to be small.  Thus, we will
not consider the limit $K\rightarrow \infty$.  We will. nonetheless, be
able to explore a large region of the phase diagram.
\par
\begin{figure}
\centerline{\psfig{figure=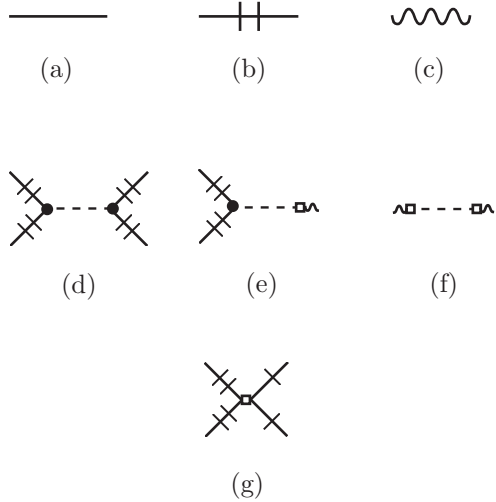}}
\caption{Elements for constructing a diagrammtic perturbation expansion for
a coulomb gas on a fluctuating membrane: (a) The height propagator $G_{hh}
= 1/(\beta \kappa q^4)$, (b) the height propagator with two gradients
represented by the vertical lines, (c) the bare vortex charge-density
propagator $C^{0}_{\tilde{n}\tilde{n}}$ propagator 
[Eq. (\protect\ref{cnn-low-y})], (d) the
coulomb vertex $K^{\prime}/q^2$ in ${\cal H}_{SS}$, 
(e) the coulomb vertex $-K/q^2$
in ${\cal H}_{nS}$, (f) the coulomb vertex 
$K/q^2$ in ${\cal H}_{nn}$, and (g) a nonlinear
vertex from the curvature energy [Eq. (\protect\ref{non-lin-cur})].}
\label{diagrams}
\end{figure}
\begin{figure}
\centerline{\psfig{figure=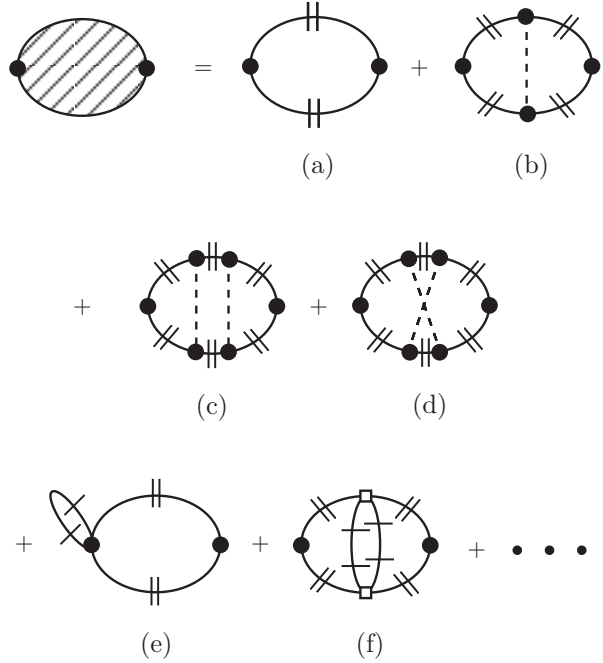}}
\caption{Diagrammatic contributions to the polarization bubble ${\cal P}$ 
represented by the cross-hatched diagram to the left of the equal sign.
The components of these diagrams are defined in Fig. 1.
The leading order contribution to  
$\cal P$ is given by diagram (a).  Diagram (e) is the first correction to
$\cal P$ arising from the ${\protect\sqrt{g}}$ factors in the definition of 
$\cal P$. The double crossline on the height propagators represent the
gradients, with appropriate symmetry, in the Gaussian curvature [Eq.
(\protect\ref{gau-cur-mon})].}
\label{coulomb}
\end{figure}
\begin{figure}
\centerline{\psfig{figure=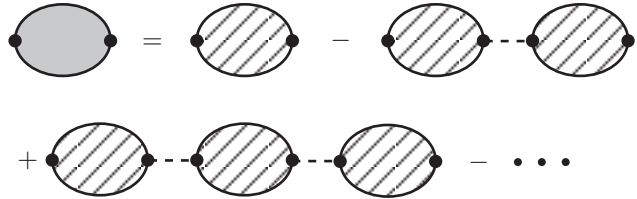}}
\caption{Diagrammatic representation of the full $C_{{\tilde n}{\tilde n}}^0$
propagator when ${\cal H}_{nS} = 0$.}
\label{Cnn}
\end{figure}

The evaluation of $C_{\tilde{\rho}\tilde{\rho}}({\bf q})$ is obtained
most directly in terms of diagrams. Figure \ref{diagrams} shows the basic
ingredients of a diagrammatic perturbation theory: the height propagator
$G_{hh}({\bf q})=(\beta\kappa q^{4})^{-1}$, the three Coulomb interactions
[Eq.~(\ref{eq:tot-c-h})] $K'/q^{2}$, $K/q^{2}$, 
and $K/q^{2}$ coupling, respectively,
$\tilde{S}$ to $\tilde{S}$, $\tilde{n}$ to $\tilde{n}$ and
$\tilde{S}$ to $\tilde{n}$, the $\tilde{n}-\tilde{n}$ propagator when
${\cal H}_{nS}$ is zero, and non-linear vertices arising from
the curvature. 
The total Coulomb Hamiltonian of Eq.~\ref{eq:tot-c-h} contains $S-S$ and
$n-n$ terms and a part
\begin{equation}
{\cal H}_{nS} = - K \int d^{2}u\sqrt{g} n\left(-\frac{1}{\Delta_{g}}
\right) \sqrt{g}S
\end{equation}
coupling $n$ to $S$.
We will first calculate correlation functions when ${\cal H}_{nS}$ is zero.
Then we will calculate the effects of turning on ${\cal H}_{nS}$.
If ${\cal H}_{nS}$ is zero, 
then $C_{\tilde{n}\tilde{S}}$ where $\tilde{S} = \sqrt{g} S$ is zero, and
the Gaussian curvature correlation function $C_{\tilde{S}\tilde{S}}$ 
is simply that of a Coulomb gas with coupling $K'$.
It can be represented as a sum of polarization bubbles shown in 
Fig.\ \ref{coulomb} linked by a Coulomb 
interaction as shown in Fig.\ \ref{Cnn}.
The result is
\begin{equation}
C^{0}_{\tilde{S}\tilde{S}}({\bf q}) = \frac{{\cal P}({\bf q})}
{1+(K'/q^{2}){\cal P}({\bf q})}.
\end{equation}
To lowest order in $h$, the Gaussian curvature is
\begin{equation}
S = \frac{1}{2}(\nabla^{2}h\nabla^{2}h - 
\partial_{i}\partial_{j}h\partial_{i}\partial_{j}h),
\label{gau-cur-mon}
\end{equation}
and, to lowest order in $(\beta\kappa)^{-1}$, ${\cal P}({\bf q})$ is 
determined by 
a diagram (a) in Fig.\ \ref{coulomb}:
\begin{eqnarray}
{\cal P}({\bf q}) & = & \frac{1}{2}\frac{1}{(\beta\kappa)^{2}}
                    \int\frac{d^{2}k}{(2\pi)^{2}}
            \frac{({\bf q} \times {\bf k})^{4}}{k^{4}({\bf k}+{\bf q})^{4}}
            \nonumber   \\
              & = & \frac{3}{32\pi} \frac{q^{2}}{(\beta\kappa)^{-2}}.
\label{eq:p-polar}
\end{eqnarray}
Note that ${\cal P}({\bf q})$ is proportional to $q^{2}$ even though 
the numerator in
the integrand is proportional to $q^{4}$. There is an infrared singularity in
the integral over ${\bf k}$ that converts the $q^{4}$ to the $q^{2}$.
Higher order contributions to ${\cal P}$ are shown in Figs.\ \ref{coulomb}
(b)-(f).
The contribution to
$\cal P$ from Fig. \ref{coulomb}b is of order $q^2\beta K/(\beta\kappa)^4$,
i.e., of order $\beta K/(\beta \kappa)^2$ smaller than Eq.\
(\ref{eq:p-polar}).  The diagrams of Fig.\ \ref{coulomb}c and
\ref{coulomb}d are smaller by another factor of $\beta K /(\beta
\kappa)^2$.  The diagram in Fig.\ \ref{coulomb}e, which arises from the
$\sqrt{g}$ factor and nonlinear terms in $\tilde S$,
is of order $(\beta \kappa)^{-1}$ times Eq.\ (\ref{eq:p-polar}).  
It, however, has an infrared divergence, which would have to be handled
with care if we wished to calculate to the next order in $T$. The diagram
in Fig.\ \ref{coulomb}f is of order $(\beta \kappa)^{-4}$. 
\par
\begin{figure}
\centerline{\psfig{figure=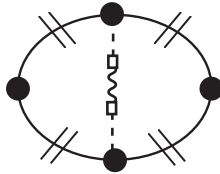}}
\caption{A representative additional diagram contibuting to ${\cal P}$ when
${\cal H}_{nS}$ is turned on.  The expansion for $C_{{\tilde n}{\tilde
n}}^0$ is still given by Eq. (\protect\ref{cnn-low-y}) and 
Fig. \protect\ref{Cnn}.}
\label{newP}
\end{figure}
\begin{figure}
\centerline{\psfig{figure=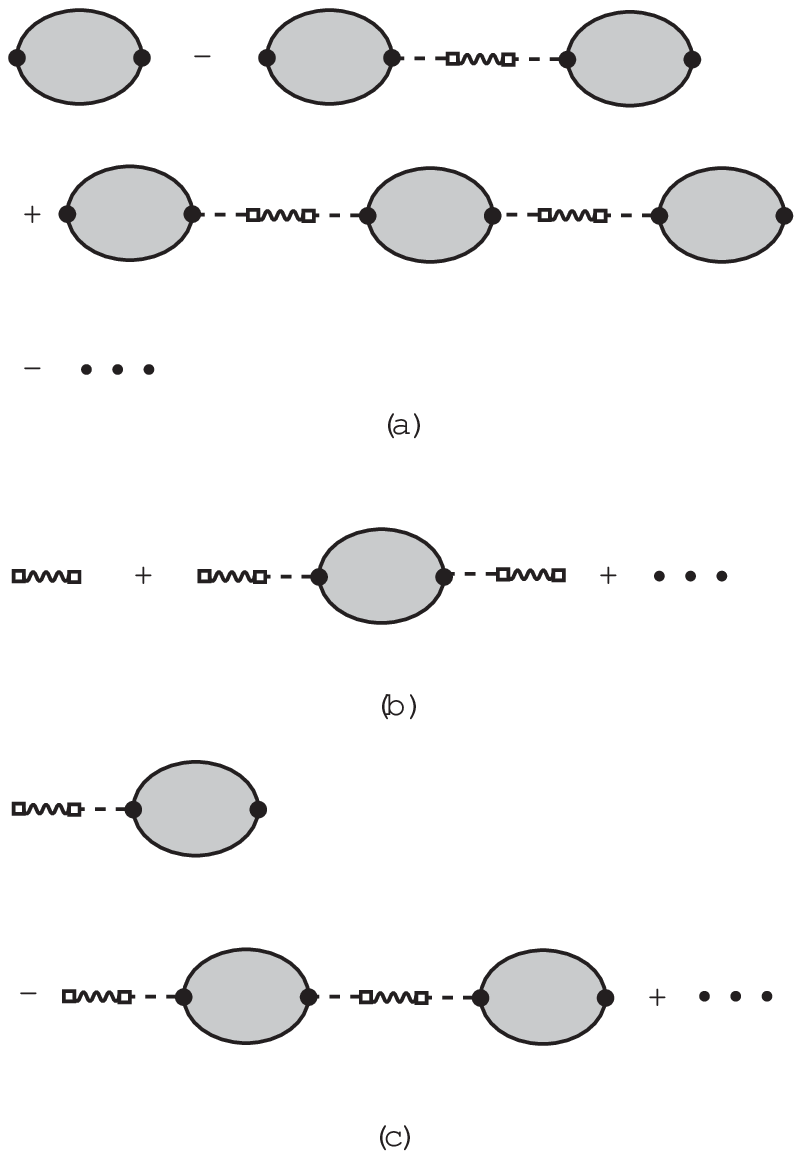}}
\caption{Diagrammatic expansion of (a) $C_{{\tilde n}{\tilde n}}$, (b)
$C_{{\tilde S}{\tilde S}}$, and (c) $C_{{\tilde n}{\tilde S}}$. 
The bubble is $C_{{\tilde n}{\tilde n}}^0$ represented in 
Fig. \protect\ref{Cnn}.}
\label{diagram2}
\end{figure}

When ${\cal H}_{nS} = 0$, the disclination correlation function
$C_{\tilde{n}\tilde{n}}({\bf q})$ can be calculated as a power series in $y$.
To lowest order in $T$, we can replace $\sqrt{g}$ by 1 in $\tilde{n}$.
Then to lowest order in $y$,
\begin{equation}
C^{0}_{\tilde{n}\tilde{n}}({\bf q}) = \frac{1}{A_{B}}\frac{1}{{\cal Z}}y^{2}
\int {\cal D}{\bf R}\int\frac{d^{2}x_{+}}{a^{2}}\frac{d^{2}x_{-}}{a^{2}}
\langle n({\bf q})n(-{\bf q})\rangle 
e^{-\beta{\cal H}_{\kappa}-\beta{\cal H}_{SS}
-\beta{\cal H}_{nn}},
\label{cnn-low-y}
\end{equation}
where $n({\bf q}) = (2\pi/p)(e^{i{\bf q}\cdot{\bf x}_{+}}-
e^{-i{\bf q}\cdot{\bf x}_{-}})$.
Therefore
\begin{eqnarray}
C^{0}_{\tilde{n}\tilde{n}}({\bf q}) & = & y^{2}\left(\frac{2\pi}{p}\right)^{2}
\frac{1}{2}q^{2}\int_{a}^{\infty} d^{2}x |{\bf x}|^{2-2\pi\beta K} \nonumber \\
   & = & \frac{4\pi^{3}}{p^{2}}y^{2}q^{2}\int_{a}^{\infty} dr 
r^{3-2\pi\beta K} + {\cal O}(y^{4}).
\label{eq:cnn-form}
\end{eqnarray}
When ${\cal H}_{nS}$ is turned on, there will be additional contributions
to the polarization bubble such as shown in Fig. \ref{newP}.
The correlation functions $C_{\tilde{n}\tilde{n}}$, $C_{\tilde{S}\tilde{S}}$ 
and $C_{\tilde{n}\tilde{S}}$ can be obtained from the diagrams in Fig.
\ref{diagram2}: 
\begin{eqnarray}
C_{\tilde{n}\tilde{n}} & = & \frac{C^{0}_{\tilde{n}\tilde{n}}}
{1- (\beta^{2}K^{2}/q^{4})C^{0}_{\tilde{n}\tilde{n}}
C^{0}_{\tilde{S}\tilde{S}}}   \\
C_{\tilde{S}\tilde{S}} & = & \frac{C^{0}_{\tilde{S}\tilde{S}}}
{1- (\beta^{2}K^{2}/q^{4})C^{0}_{\tilde{n}\tilde{n}}
C^{0}_{\tilde{S}\tilde{S}}}   \nonumber  \\
   & = & \frac{{\cal P}}{1+(\beta K'/q^{2}) {\cal P} - \beta^{2}K^{2}
{\cal P}C^{0}_{\tilde{n}\tilde{n}}}  
\label{eq:c-nnss}    \\
C_{\tilde{n}\tilde{S}} & = & \frac{\beta K}{q^{2}} 
\frac{C^{0}_{\tilde{n}\tilde{n}}C^{0}_{\tilde{S}\tilde{S}}}
{1- (\beta^{2}K^{2}/q^{4})
C^{0}_{\tilde{n}\tilde{n}}C^{0}_{\tilde{S}\tilde{S}}},
\label{eq:c-ns}
\end{eqnarray}
where $C^{0}_{\tilde{S}\tilde{S}}$ is given by Eq. (\ref{eq:p-polar})
with ${\cal P}({\bf q})$ corrected by contribution such as that in 
Fig. \ref{newP}.
Combining Eqs.~(\ref{eq:c-nnss}) and~(\ref{eq:c-ns}), we obtain the 
charge-density correlation function determining $K_{R}$:
\begin{eqnarray}
C_{\tilde{\rho}\tilde{\rho}}({\bf q}) & = & C_{\tilde{n}\tilde{n}}({\bf q}) 
- 2C_{\tilde{n}\tilde{S}}({\bf q}) + C_{\tilde{S}\tilde{S}}({\bf q}) 
\nonumber  \\
   & = & \frac{C^{0}_{\tilde{n}\tilde{n}}({\bf q}) -2(\beta K/q^{2})
C^{0}_{\tilde{n}\tilde{n}}({\bf q})C_{\tilde{S}\tilde{S}}({\bf q}) +
C_{\tilde{S}\tilde{S}}({\bf q})}
{1- (\beta^{2} K^{2}/q^{4})
C^{0}_{\tilde{n}\tilde{n}}({\bf q})C_{\tilde{S}\tilde{S}}({\bf q})}  
\nonumber  \\
   & = & \frac{{\cal P}({\bf q}) + 
C^{0}_{\tilde{n}\tilde{n}}({\bf q})[1- (\beta(2K-K')/q^{2}){\cal P}({\bf q})]}
{1+ (\beta K/q^{2}) {\cal P}({\bf q})[(K'/K) - (\beta\kappa/q^{2})
C^{0}_{\tilde{n}\tilde{n}}({\bf q})]}.
\end{eqnarray}
In the limit of low-temperature and low-fugacity, this reduces to
\begin{equation}
C_{\tilde{\rho}\tilde{\rho}}({\bf q}) \simeq {\cal P}({\bf q}) + 
C^{0}_{\tilde{n}\tilde{n}}({\bf q}) \left(1-2\frac{\beta K}{q^{2}}
{\cal P}({\bf q})
\right) + \cdots,
\label{crr-ap}
\end{equation}
where ${\cal P}({\bf q})$ is given by Eq.~(\ref{eq:p-polar}) and 
$C^{0}_{\tilde{n}\tilde{n}}({\bf q})$ by Eq.~(\ref{eq:cnn-form}).
The second term in the coefficient of $C_{{\tilde n}{\tilde n}}^0$ is of
order $\beta K/(\beta \kappa)^2$ and will be neglected.
\par
We can now calculate the dielectric constant or renormalized rigidity
to lowest order in temperature and fugacity using Eqs. (\ref{x2rr}),
(\ref{eq:k-ren}), (\ref{eq:p-polar}), (\ref{eq:cnn-form}) and 
(\ref{crr-ap}).
\begin{equation}
\beta K_{R} = \beta{\overline K} - 
(\beta{\overline K})^{2}\frac{4\pi^{3}}{p^{2}}
y^{2} \int_{a}^{\infty} dr r^{3-2\pi\beta{\overline K}},
\label{eq:rge-K}
\end{equation}
where
\begin{equation}
\beta{\overline K} = \beta K - 
\frac{3}{32\pi} \left( \frac{K}{\kappa} \right)^{2}.
\label{eq:K-kappa}
\end{equation}
we assume $(3/32\pi )\beta K/(\beta \kappa)^2 \ll 1$ so that 
\begin{equation}
(\beta {\overline K})^{-1} = (\beta K)^{-1} + {3 \over 32 \pi} 
(\beta\kappa)^{-2} .
\end{equation}
Thus, the equation for $\beta K_{R}$ is identical to that for a rigid flat
membrane but with ${\overline K}$ replacing $K$.
\setcounter{equation}{0}
\section{Renormalization Equations}
To determine the properties of the Kosterlitz-Thouless transition on a
fluctuating membrane, we need to determine the renormalization flow
equations for $\kappa$, $K$, and $y$.  Equation (\ref{eq:rge-K}) for $K_R$ is
identical to the equation for $K_R$ on a flat membrane\cite{KT,jp-lub} with $K$
replace by $\overline K$.  Flow equations for ${\overline K}(l)$ and $y(l)$
can be obtained in the usual way\cite{Jose} 
from Eq. (\ref{eq:rge-K}). By breaking up the 
integral $\int_a^\infty dr$ into 
$\int_a^{ae^l} dr + \int_{ae^l}^\infty dr$, putting
the first part into a renormalized ${\overline K}(l)$ and rescaling the
second part by $r \rightarrow r e^{-l}$, we obtain
\begin{equation}
(\beta K_R )^{-1} = (\beta {\overline K}(l) )^{-1} +{4 \pi^3 \over p^2}
y^2(l) \int_a^{\infty} dr r^{3 - 2 \pi \beta {\overline K}}
\end{equation}
with
\begin{eqnarray}
& &{d (\beta{\overline K})^{-1}\over dl} =  {4 \pi^3 \over p^2} y^2 
\label{RG1}\\
& &{d y \over dl}  =  \left( 2 - {\pi \beta {\overline K} \over p^2} \right)
y .
\label{RG2}
\end{eqnarray}
Observe that the bending rigidity does not appear explicitly in these
equations; it only appears via the dependence of $\beta {\overline K}$ on
$\kappa$.  This suggests that there should be an effective theory for a
flat membrane with a bare hexatic rigidity equal to ${\overline K} = K - (3
/32 \pi) (K/\kappa)^2$ rather than $K$.  Indeed, if we integrate our height
fluctuations to produce an effective flat space Hamiltonian via
\begin{equation}
e^{-\beta{\cal H}_{\rm eff}} = \int {\cal D} h e^{- \beta{\cal H}_{\kappa} 
- \beta{\cal H}_{\theta}} ,
\end{equation}
we obtain
\begin{equation}
{\cal H}_{\rm eff} = \case{1}{2} \beta {\overline K} \int d^2 x 
( \nabla \theta )^2 .
\label{Heff}
\end{equation}
Thus, the fluctuating membrane behaves like a flat membrane with a hexatic
rigidity ${\overline K}$.  Note that the renormalized rigidity $\overline
K$ appears in Eq.\ (\ref{Heff}) because fluctuations at {\em all}
wavenumbers ${\bf q}$ have been integrated out.  If we had calculated an
effective Hamiltonian by removing $h( {\bf q} )$ for ${\bf q}$ within a
shell $\Lambda / b < q < \Lambda$, there would have been no shift in $K$.
\par
\begin{figure}
\centerline{\psfig{figure=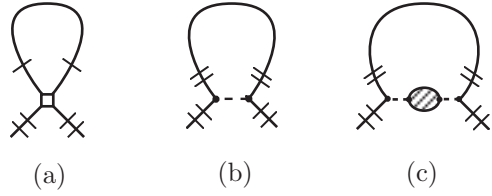}}
\caption{Diagrams for the bending rigidity $\kappa$}
\label{diagram1}
\end{figure}
\begin{figure}
\centerline{\psfig{figure=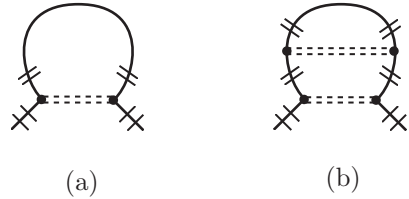}}
\caption{Diagrams contributing to $\kappa$ in terms of the screened
Coulomb propagator (represented by the double dashed line)
$K^{\prime}/(\epsilon_{\parallel,SS} q^2)$ defined in 
Eq. (\protect\ref{epsilon-para}).}
\label{sceenkappa}
\end{figure}

The renormalization of $\kappa$ can be calculated using 
Eqs. (\ref{eq:tr-hs}) and (\ref{eq:tot-c-h}).
Diagrams contributing to the renormalized bending rigidity $\kappa_R$ to
lowest order in $\beta^{-1}$ are shown in Fig.\ \ref{diagram1}.  They yield
\begin{eqnarray}
\beta \kappa_R & = & \beta \kappa 
   - {3 \over 4 \pi} \int_{L^{-1}}^{\Lambda} {d q \over q} +
   {3 \over 4 \pi} {\beta K' \over 4 \beta \kappa} \int_{L^{-1}}^{\Lambda}
   {d q \over q} 
   + \frac{3}{4\pi}\frac{\pi^{3}y^{2}}{p^{2}}\frac{(\beta K')^{2}}
   {4\beta\kappa} \left( \int_{L^{-1}}^{\Lambda}\frac{dq}{q} \right)^{2}
   \nonumber \\
& = & \beta \kappa -{3 \over 4 \pi} \left(1 - {\beta K' \over 4 \beta \kappa}
   \right) \int_a^L{dr \over r} 
   +\frac{3}{4\pi}\frac{\pi^{3}y^{2}}{p^{2}}\frac{(\beta K')^{2}}
   {4\beta\kappa} \left( \int_{a}^{L}\frac{dr}{r} \right)^{2},
\end{eqnarray}
where $L$ is the linear dimension of the membrane,
which we take to be infinite, and $\Lambda = 1/a$.  We
can now follow exactly the same procedure to obtain the recursion relation
for $\kappa$ we used to obtain those for ${\overline K}$ and $y$. 
To lowest order in $(\beta\kappa)^{-1}$, we can replace $K'$ by $K$.
We break
the integral $\int_a^{\infty}$ into two parts, and rescale to obtain
\begin{equation}
{d \beta\kappa \over dl} = -{3 \over 4 \pi} \left(1 - {\beta K\over 4 \beta
\kappa}\right) .
\label{kappa1}
\end{equation}
This equation was derived from the low-order diagrams shown in 
Fig.\ \ref{diagram1} in
which the Coulomb energy $K/q^2$ is treated as a perturbation.  One can
also structure a perturbation theory in which the Coulomb propagator
$K'/q^{2}$ coupling $\tilde{S}$ to $\tilde{S}$ [Eq. (\ref{eq:tot-c-h})] is
replaced by $K'/\epsilon_{\parallel,\tilde{S}\tilde{S}}q^2$ where
\begin{equation}
\epsilon_{\parallel,\tilde{S}\tilde{S}}^{-1}({\bf q}) = 1-\frac{K'}{q^{2}}
C_{\tilde{S}\tilde{S}}({\bf q}) - 
\frac{K}{K'}\frac{1}{q^{2}}C_{\tilde{n}\tilde{n}}
+ \frac{2K}{q^{2}}C_{\tilde{n}\tilde{S}}
\label{epsilon-para}
\end{equation}
as shown in Fig.\ \ref{sceenkappa}.  
Similar expressions can be derived for the effective Coulomb potentials
coupling $\tilde{n}$ to $\tilde{n}$ and $\tilde{n}$ to $\tilde{S}$.
In this scheme, 
Figs.\ \ref{diagram1}b and c are replaced by Fig.\ \ref{sceenkappa}a.  
In addition, there are higher order diagrams such as
those shown in Fig.\ \ref{sceenkappa}b.  
These diagrams have two or more loops and are
difficult to treat using our naive renormalization scheme.  In the
companion paper, we will employ a controlled regularization scheme to treat
these graphs using the sine-Gordon Hamiltonian.  It shows that $K$ should be
replaced by ${\overline K}$ in Eq. (\ref{kappa1}).  Thus we have
\begin{equation}
{d \beta \kappa \over dl} = -{3 \over 4 \pi} \left( 1 - {\beta {\overline
K}\over 4 \beta \kappa} \right) .
\label{RG3}
\end{equation}
The renormalization flow equations now depend on ${\overline K}$ and
not on $K$.
\par
\begin{figure}
\centerline{\psfig{figure=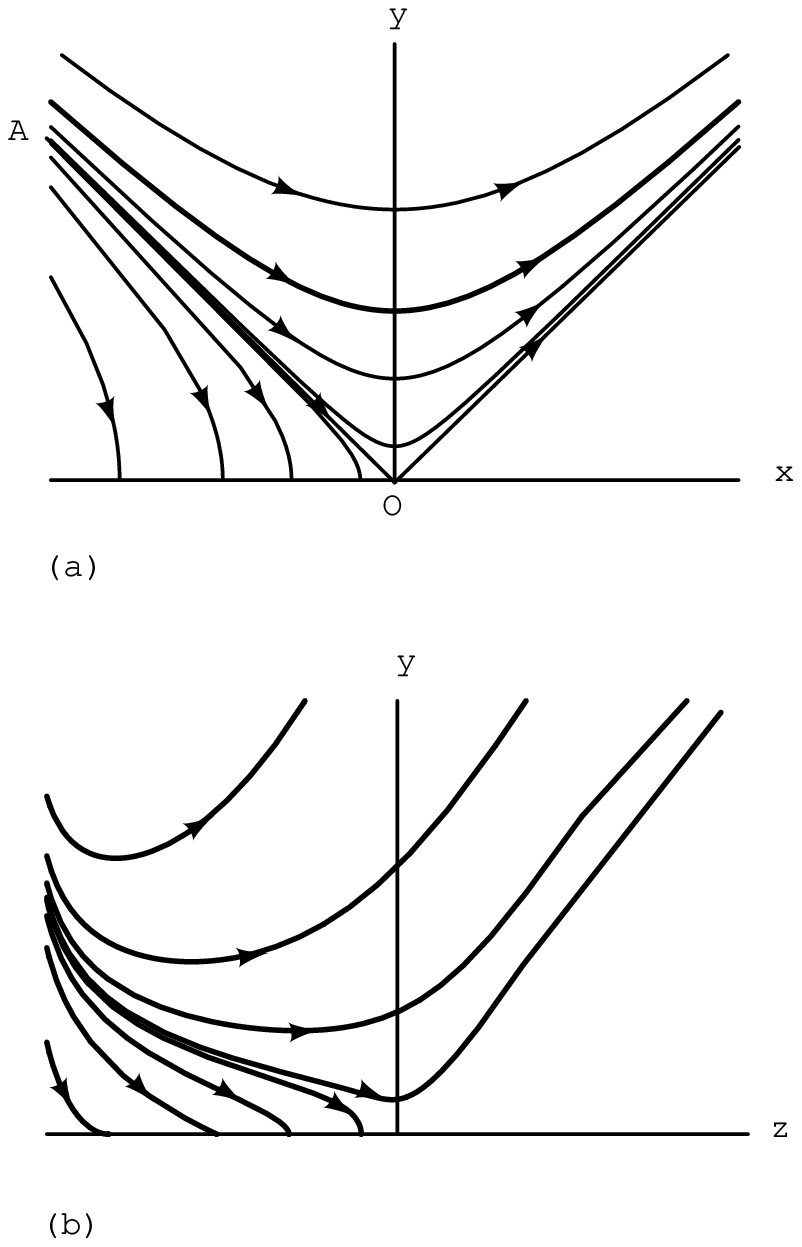}}
\caption{(a) Renormalization flows in the $xy$-plane.  These are identical
to the order of our calculations to those of the flat space $xy$-model.
The separatrix $AO$ is the critical line.
(b) Flows in the $zy$-plane.  They are similar to those in the $xy$-plane.}
\label{fig1}
\end{figure}
\begin{figure}
\centerline{\psfig{figure=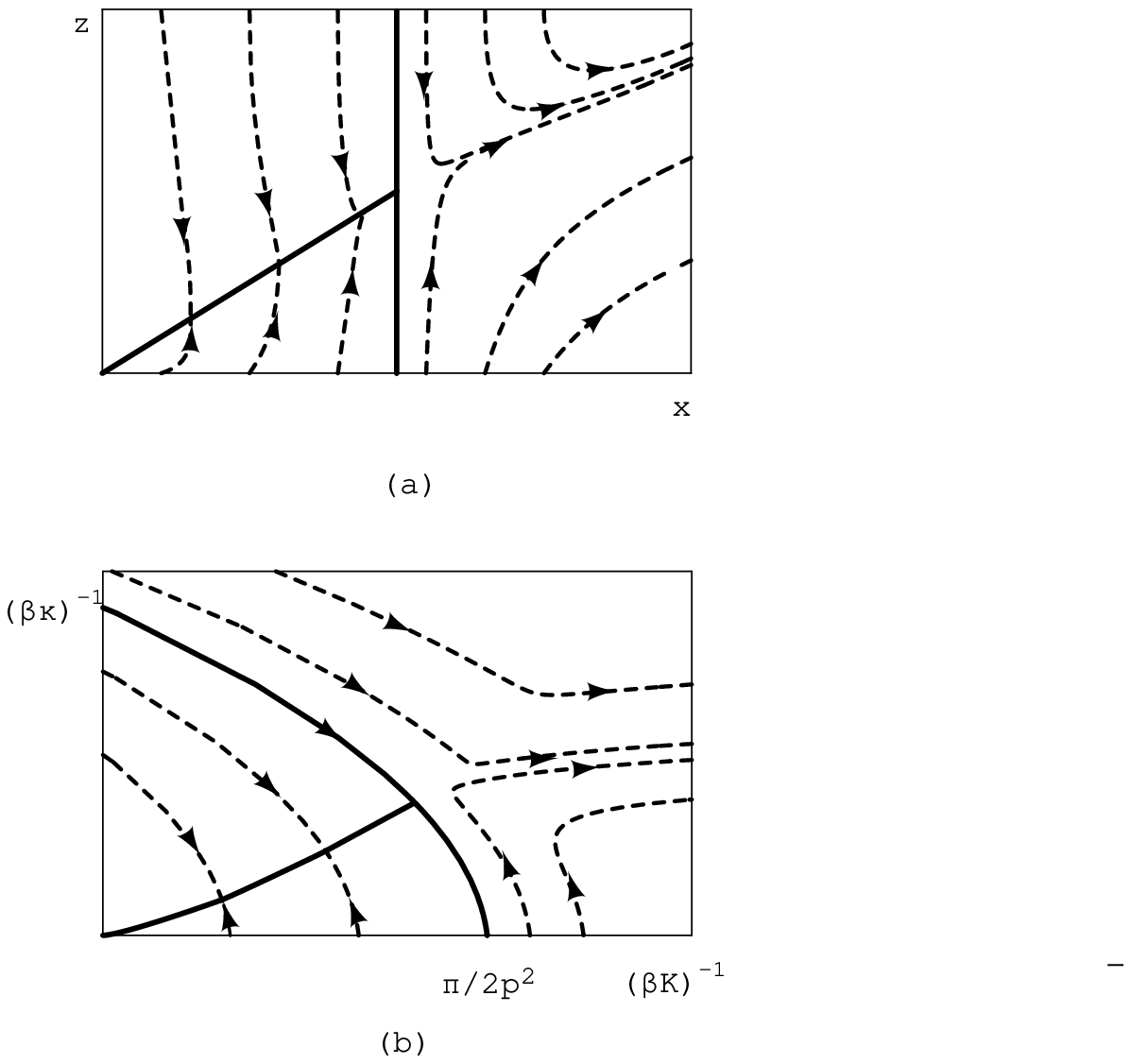}}
\caption{(a) Renormalization flows in the $xz$ plane.  The transition from
the crinkled to the crumpled phase takes place at $x=0$ independent of $z$
(i.e., independent of $\beta \kappa$),
in agreement with previous calculations\protect\cite{GuitKardar90-1}. 
(b) Flows in
the $(\beta K)^{-1}$-$(\beta \kappa)^{-1}$ plane.  Here, there is a
transition from the crinkled to the crumpled phase as $\kappa$ is
decreased.}   
\label{fig2}
\end{figure}

Equations (\ref{RG1}), (\ref{RG2}), and (\ref{RG3}) define renormalization
flows for fluctuating hexatic membranes.  They have a fixed point at 
\begin{equation}
\beta{\overline K}^* = {2 p^2\over \pi} , \qquad 
y^* = 0 , \qquad \beta \kappa^* = \beta
{{\overline K}^* \over 4} = {p^2 \over 2 \pi} .
\end{equation}
We can obtain equations in the vicinity of this fixed point by defining
\begin{equation}
\beta{\overline K}= \beta{\overline K}^* ( 1 - x ) ,\qquad
\beta\kappa = \beta \kappa^* ( 1 - z) .
\label{lineariz}
\end{equation}   
Then
\begin{equation}
{d x \over dl } =  8 \pi^2 y ^2 , \qquad
{d y \over dl} = 2 xy
\label{nrecur1}
\end{equation}
\begin{equation}
{d z \over dl }=  {3 \over 2 p^2 } { x - z \over 1-z}.
\label{nrecur3}
\end{equation}
The flow lines for these equations are shown in Figs.\ \ref{fig1} 
and Fig.\ \ref{fig2}.  Figure \ref{fig1}a shows flows in the $xy$-plane,
which are identical to those in flat space \cite{KT}.  Figure \ref{fig1}b
shows flows in the $yz$-plane, which are similar to those in the
$xy$-plane.
Figure \ref{fig2}a shows flows in the $(\beta {\overline K})^{-1}$-$(\beta
\kappa )^{-1}$ plane.  As in previous treatments\cite{David87-1}, there is a
fixed line in this plane at $\beta \kappa = \beta {\overline K}/4$ ($x=z$),
and the crinkled-to-crumpled transition occurs at $\beta {\overline K} = 
\beta {\overline K}^*$ independent of $\beta \kappa$, and the
Kosterlitz-Thouless transition occurs at $x=0$ independent of $z$.  
\par 
The separatrix $AO$ in Fig.\ \ref{fig1}a is the transition line dividing the
crinkled hexatic phase from the melted crumpled phase.  
Systems whose initial values
of $x$ and $y$ lie below and to the left of the separatrix flow to the
low-temperature crinkled hexatic phase; points above and to the right of the
separatrix flow away to the crumpled fluid.  The initial value of $x$ is 
$x_0 = 1 - {\overline K}/{\overline K}^*$, which increases with decreasing
$\beta {\overline K} = \beta K - (3/32 \pi) (K/\kappa)^2$.  Decreasing the
bending rigidity $\kappa$ decreases ${\overline K}$ and eventually drives
$x$ to the right of  the separatrix.  Thus, there will be a transition 
from the
crinkled to the crumpled phase as $\kappa$ is decreased.  An alternative way
to see this is to use $K$ rather than $\overline K$ as the independent
variable.  To do this, we define $K(l)$ to satisfy 
Eq.\ (\ref{eq:K-kappa}) with
${\overline K}$ and $\kappa$ replaced, respectively, by ${\overline K} (l)$
and $\kappa (l)$. Flows in the $((\beta K(l))^{-1}, (\beta \kappa ( l )
)^{-1} )$ plane are shown in Fig.\ \ref{fig2}b.  The vertical transition
line in the $((\beta {\overline K} (l))^{-1}, (\beta \kappa (l))^{-1} )$
plane is now curved and crosses the $(\beta K)^{-1} = 0$ 
axis at a finite value
of $(\beta \kappa )^{-1}$.  Points to the left and below the separatrix
flow to the crinkled fixed line.  
Points outside it flow to the crumpled phase.
\par
In deriving our recursions relations, we restricted ourselves to $(\beta
\kappa )^{-1}$, $(\beta K)^{-1}$ and $\beta K/(\beta \kappa )^2 < 2 \pi$.
Thus, our calculations strictly speaking only apply below the curves $OA$
in Figs.\ \ref{hexphased} and \ref{fig2}b.  The flows in Fig.\ \ref{fig2}
we calculated for $p=1$.  Figure \ref{hexphased} shows flows for the
physically more interesting case of $p=6$.  Our approximation applies to
a considerable region around the fixed point $P$.  Higher order terms in
$(\beta \kappa)^{-1}$ and $\beta K/(\beta \kappa )^2$ would have to be
included to get an accurate picture of what happens above the curve $OA$.
Is seems likely to us, however, that these higher order terms will not
lead to any qualitative modifications to Fig.\ \ref{hexphased}.  For any
finite value of $K^{-1}$ and $\kappa^{-1}$, the crinkled-to-crumpled
transition is governed by the fixed point $P$.  For infinite hexatic
rigidity, $K^{-1} = 0$, the transition is controlled by different physics,
probably analogous to that of tethered membranes\cite{tethered}.  There
could, of course, be some phase boundary at large $\beta K/(\beta
\kappa)^2$ separating $K = \infty$-like behavior from finite-$K$
behavior, but we do not see any particular reason why this should be so.
\par
The persistence length in the fluid phase is
\begin{equation}
\xi_p = a e^{l^*} ,
\end{equation}
where $l^*$ is determined by 
\begin{equation}
\beta \kappa ( l^* ) = \beta \kappa^* ( 1 - z ( l^* )) = 0 ,
\end{equation}
i.e., by $z(l^* ) = 1$.  Integration of Eqs. (\ref{nrecur1}) and 
(\ref{nrecur3}) yield $z(l)$ and thus 
$\xi_p$.  Equation (\ref{nrecur3}) for $z$ is a nonlinear and 
cannot be determined
analytically.  We can, however, obtain a very good estimate of $\xi_p$ when
$\xi_p > \xi_{KT}$, where $\xi_{KT} \sim a \exp( b/|T - T_{KT}|^{1/2})$ 
is the Kosterlitz-Thouless correlation length.  The latter length is
$\xi_{KT} = a e^{l_0}$, where $\beta {\overline K} (l_0 ) = 0$ or $x(l_0 )
= 1$.  The recursion relation, Eq.\ \ref{RG3} for $z(l)$ is only valid for
${\overline K} > 0$.  When ${\overline K} \rightarrow 0$, we approach a
fluid phase with recursion relations for $\kappa$ determine by $\kappa$
alone.  We, therefore, assume that $x(l) = 0$ for $l > l_0$.  Then
\begin{equation}
{d z \over dl} = {3 \over 2 p^2} , \qquad \qquad l > l_0 .
\end{equation}
This equation can be integrated subject to the boundary condition that
$z(l_0)$ be the value of $z(l)$ at $l= l_0$ determined by 
Eqs.\ \ref{RG1} and \ref{lineariz}.
Thus,
\begin{equation}
z(l ) = z(l_0) + {3 \over 2 p^2} ( l - l_0 ) ,
\end{equation}
$l^* = l_0 + 2 p^2 (1 - z(l_0 )) /3 $, and
\begin{equation}
\xi_p = a e^{l^*} = \xi_{KT} e^{4 \pi \beta \kappa ( l_0 )/3} ,
\label{xip}
\end{equation}
where we used $\beta \kappa ( l_0 ) = p^2 ( 1 - z(l_0 ) / ( 2 \pi )$.
This expression is to be compared with the result $\xi_p = a e^{4 \pi \beta
\kappa /3} $ for a pure fluid membrane.  It is what one would naively have
expected.  The microscopic length $a$ is replaced by the KT coherence
length $\xi_{KT}$, and $\kappa$ at length scale $a$ is replace by $\kappa$
at length scale $\xi_{KT}$.  Equation \ref{xip} for $\xi_p$ is valid
provided $\xi_p > \xi_{KT}$ (or $l^* > l_0$).  Near the KT critical point,
this inequality may not be satisfied.
\setcounter{equation}{0}
\section{Review and Discussion}
\par
We have investigated the Kosterlitz-Thouless transition on fluctuating
hexatic membranes.  We developed three equivalent Hamiltonians for
describing these membranes: the hexatic elastic Hamiltonian expressed in
terms of gradients of the hexatic angle variable, a Coulomb-gas
Hamiltonian, and a sine-Gordon Hamiltonian.  The Coulomb gas is
characterized by Gaussian-curvature and disclination charge densities,
which interact via potentials partially determined by the Liouville action
arising from a covariant cutoff.  The sine-Gordon Hamiltonian has a linear
coupling between the sine-Gordon field and Gaussian curvature.  We showed
that height fluctuations, when integrated over all wave number, soften the
hexatic stiffness $K$.  As a result, the 
disclination melting transition from the hexatic
crinkled phase to the fluid crumpled phase is brought about both by
increasing temperature and by decreasing the bending rigidity $\kappa$.
We derived renormalization group recursion relations for $K$, $\kappa$,
and the disclination fugacity $y$ that explicitly verify this.  In a
companion paper, we provide an alternative derivation of these recursion
relations using the sine-Gordon Hamiltonian.
\par
Though the picture we present of the Kosterlitz-Thouless transition from
the hexatic crinkled phase to the crumpled fluid phase makes a great deal
of sense, it does leave some incompletely answered or unanswered questions.
First, we believe that the nature of shape fluctuation in the crinkled is
not completely resolved.  It is generally believed\cite{David87-1} that the
crinkled phase is characterized by power-law correlations in layer
normals:
$\langle{\bf N}({\bf x})\cdot {\bf N} ( 0 ) \rangle \sim |{\bf x}|^{-
\eta}$, where $\eta = 2T/( \pi K)$.  This result is obtained via
exponentiation of the expansion
\begin{eqnarray}
\langle {\bf N} ( {\bf x} ) \cdot {\bf N} ( 0 ) \rangle & \approx &
1 - \case{1}{2} \langle [ \nabla h ( {\bf x} - \nabla h ( 0 ) ]^2 \rangle
\nonumber \\
& & = 1 - {T \over 2 \pi \kappa} \ln (| {\bf x} | /a ) 
\label{NN}
\end{eqnarray}
using $\kappa = K/4$.  The exponentiation of this series has not been
explicitly justified with, for example, a calculation of second order
terms in $(T/\kappa) \ln ({\bf x} /a )$.  It is interesting to note that
the de Gennes-Taupin persistence length\cite{deGTau} 
$\xi_{GT} = a e^{2 \pi \kappa /
T}$ beyond which a fluid membrane is crumpled was calculated using Eq.\
(\ref{NN}) and setting $\langle {\bf N} ({\bf x}) \cdot {\bf N} ( 0 )
\rangle = 0$.  This observation would suggest that the possibility that
the crinkled phase is in fact crumpled, but with a longer persistence
length than the fluid phase, cannot be ruled out.
\par
Second, we have in our calculations imposed a constraint of charge
neutrality. In flat membranes (i.e., films), this constraint is imposed
by the prohibitive energy cost of having excess charge of either sign.
We believe that there is a similar energy cost for breaking charge
neutrality in free membranes, which are not constrained to be flat,
though the situation here is more complicated.  Nelson\cite{NEL} has
pointed out that the plus-minus symmetry present in flat membranes is
broken in free membranes.  Membranes with a single disclination undergo a
mechanical buckling transition from a flat configuration to a cone
configuration for a positive disclination or a saddle configuration for a
negative disclination.  The energy of the cone is lower than that of the
saddle, though both energies are proportional to $\ln (R/a)$ where $R$ is
the linear dimension of the membrane.  Does this asymmetry modify the
picture presented in this paper?  We believe not.  A free membrane will
choose configurations that will minimize its free energy.  Any
configuration, whether flat or buckled, with an excess of one sign of
charge will have a contribution to its energy proportional to $\ln
(R/a)$.  Charge neutral configurations, on the other hand, have energies
that are finite in the $R\rightarrow \infty$ limit.  Thus, there is a
prohibitive energy cost in both flat and free membranes to the violation
of charge neutrality.  The shape of a membrane near the core of positive
and negative disclination may nonetheless differ and lead to different
fugacities $y_+$ and $y_-$.  We generalize our treatment to include
this possibility in a companion paper.  
The results are that the ratio of the two 
fugacities is a marginal variable. In the ordered phase, both $y_+$ and
$y_-$ scale to zero, and the KT transition to the disordered phase is not
affected.  In the disordered phase, both fugacities grow with a fixed
ratio.  A more complete treatment of Gaussian curvature will be needed
to interpret this result.  
\par
We are grateful to Mark Bowich, David Nelson, Phil Nelson, Burt Ovrut, 
and Leo Radzihovski for helpful conversations.  This work was supported in
part by the Penn Laboratory for Research in the Structure of Matter under
NSF grant No.  91-20668.  Further support was provided by the Institute for
Theoretical Physics (under NSF grant No. 89-04035), where a portion of this work
was carried out.

\def\Dir{/u/tom/Papers/ktmem}
\input psfig
\newpage
\end{document}